\documentstyle[11pt,psfig]{article}
 1
\font\elevenrm=cmr10 scaled\magstep 1
 1

\newcommand{\beqn}{\begin{eqnarray}}
\newcommand{\eeqn}{\end{eqnarray}}

\renewenvironment{thebibliography}[1]
 { \elevenrm
   \begin{list}{\arabic{enumi}.}
    {\usecounter{enumi} \setlength{\parsep}{0pt}
     \setlength{\itemsep}{3pt} \settowidth{\labelwidth}{#1.}
     \sloppy
    }}{\end{list}}
\makeatletter
\def\section{\@startsection{section}{0}{\z@}{5.5ex plus .5ex minus
  1.5ex}{2.3ex plus .2ex}{\bf}}
\renewcommand{\theequation}{\rm\thesection.\arabic{equation}}

\newcommand{\appendixA}{\setcounter{equation}{0}
  \def\theequation{\rm{A}.\arabic{equation}}\section*}
\newcommand{\appendixB}{\setcounter{equation}{0}
  \def\theequation{\rm{B}.\arabic{equation}}\section*}
\newcommand{\appendixC}{\setcounter{equation}{0}
  \def\theequation{\rm{C}.\arabic{equation}}\section*}
\newcommand{\appendixD}{\setcounter{equation}{0}
  \def\theequation{\rm{D}.\arabic{equation}}\section*}
\makeatother

\makeatletter
\newcommand{\fcaption}[1]{
        \refstepcounter{figure}
        \setbox\@tempboxa = \hbox{\small Fig.~\thefigure. #1}
        \ifdim \wd\@tempboxa > 14.4cm
           {\begin{center}
        \parbox{14.4cm}{\small Fig.~\thefigure. #1}
            \end{center}}
        \else
             {\begin{center}
             {\small Fig.~\thefigure. #1}
              \end{center}}
        \fi}

\newcount\@tempcntc
\def\@citex[#1]#2{\if@filesw\immediate\write\@auxout{\string\citation{#2}}\fi
  \@tempcnta\z@\@tempcntb\m@ne\def\@citea{}\@cite{\@for\@citeb:=#2\do
    {\@ifundefined
       {b@\@citeb}{\@citeo\@tempcntb\m@ne\@citea\def\@citea{,}{\bf ?}\@warning
       {Citation `\@citeb' on page \thepage \space undefined}}%
    {\setbox\z@\hbox{\global\@tempcntc0\csname b@\@citeb\endcsname\relax}%
     \ifnum\@tempcntc=\z@ \@citeo\@tempcntb\m@ne
       \@citea\def\@citea{,}\hbox{\csname b@\@citeb\endcsname}%
     \else
      \advance\@tempcntb\@ne
      \ifnum\@tempcntb=\@tempcntc
      \else\advance\@tempcntb\m@ne\@citeo
      \@tempcnta\@tempcntc\@tempcntb\@tempcntc\fi\fi}}\@citeo}{#1}}
\def\@citeo{\ifnum\@tempcnta>\@tempcntb\else\@citea\def\@citea{,}%
  \ifnum\@tempcnta=\@tempcntb\the\@tempcnta\else
   {\advance\@tempcnta\@ne\ifnum\@tempcnta=\@tempcntb \else \def\@citea{--}\fi
    \advance\@tempcnta\m@ne\the\@tempcnta\@citea\the\@tempcntb}\fi\fi}
\makeatother
\def\CERN{{\it CERN, TH-Division}, 
  {\it CH-1211 Geneva 23, Switzerland} \\}
\def\SCIPP{{\it Santa Cruz Institute for Particle Physics}\\
  {\it University of California, Santa Cruz, CA 95064 USA} \\}
\def\KARL{{\it Institut f\"ur Theoretische Teilchenphysik,
 Universit\"at Karlsruhe,}\\
  {\it Kaiserstrasse 12, D-76128 Karlsruhe, Germany} \\}
\def\MPI{{\it Max-Planck-Institut f\"ur Physik, Werner-Heisenberg-Institut,}\\
   {\it F\"ohringer Ring 6, D-80805 Munich, Germany}\\}
\renewcommand{\thefootnote}{\fnsymbol{footnote}}
\begin{document}
\noindent
\thispagestyle{empty}
\begin{flushright}
{\bf CERN-TH/95-216}\\
{\bf SCIPP 95/52}\\
{\bf MPI-PhT/95-103}\\
{\bf NSF-ITP-96-62} \\
{\bf TTP95-09}\footnote{The postscript file of this
preprint, including figures, is available via anonymous ftp at
ftp://www-ttp.physik.uni-karlsruhe.de (129.13.102.139) as
/ttp95-09/ttp95-09.ps or via www at 
http://www-ttp.physik.uni-karlsruhe.de/cgi-bin/preprints/.}\\
hep-ph/9609331\\
\end{flushright}
  \vspace{0.25cm}
\begin{center}
\begin{bf}
 \begin{Large}
APPROXIMATING THE RADIATIVELY CORRECTED HIGGS MASS\\
IN THE MINIMAL SUPERSYMMETRIC MODEL\\
 \end{Large}
\end{bf}
  \vspace{0.5cm}
  \begin{large}
  Howard E. Haber,$^1$
%\footnote{Permanent
%       address: Santa Cruz Institute for Particle Physics, University of
%       California, Santa Cruz, CA 95064 USA.}$^{,a}$,
  Ralf Hempfling$^2$ and Andr\'e H. Hoang$^3$\\
  \end{large}
  \vspace{0.5cm}
% \vspace{0.5cm}
  $^1$\CERN 
\vspace{0.1cm}
{and}\\
\vspace{0.1cm}
 \SCIPP
  \vspace{0.5cm}
  $^2$\MPI
  \vspace{0.5cm}
  $^3$\KARL
  \vspace{1.0cm}
  {\bf Abstract}\\
\vspace{0.5cm}
\noindent
\begin{minipage}{13.0cm}
\begin{small}
To obtain the most accurate predictions for the Higgs masses in the
minimal supersymmetric model (MSSM), one should compute the full set of
one-loop radiative corrections, resum the large logarithms to all
orders, and add the dominant two-loop effects.  A complete
computation following this procedure yields a complex set of formulae
which must be analyzed numerically.  We discuss a very simple
approximation scheme which includes the most important terms from
each of the three components mentioned above.  We estimate that the
Higgs masses computed using our scheme lie within 2~GeV of their
theoretically predicted values over a very large fraction of MSSM
parameter space.
\end{small}
\end{minipage}
\end{center}
\vfill
\vspace{0.25cm}
\vbox{\hbox{CERN-TH/95-216}
      \hbox{September, 1996}
      \hbox{\mbox{}}}
\setcounter{footnote}{0}
\renewcommand{\thefootnote}{\alph{footnote}}
\newpage
\vspace{1.2cm}
\def\lsim{\mathrel{\raise.3ex\hbox{$<$\kern-.75em\lower1ex\hbox{$\sim$}}}}
\def\gsim{\mathrel{\raise.3ex\hbox{$>$\kern-.75em\lower1ex\hbox{$\sim$}}}}
\def\ifmath#1{\relax\ifmmode #1\else $#1$\fi}
\def\half{\ifmath{{\textstyle{1 \over 2}}}}
\def\quarter{\ifmath{{\textstyle{1 \over 4}}}}
\def\3quarter{{\textstyle{3 \over 4}}}
\def\ninequarters{{\textstyle{9 \over 4}}}
\def\ninehalves{{\textstyle{9 \over 2}}}
\def\third{\ifmath{{\textstyle{1 \over 3}}}}
\def\eightthirds{\ifmath{{\textstyle{8 \over 3}}}}
\def\twothirds{{\textstyle{2 \over 3}}}
\def\fourth{\ifmath{{\textstyle{1\over 4}}}}
\def\svttwelves{\ifmath{{\textstyle{17\over 12}}}}
\def\tanb{\tan\beta}
\def\calo{{\cal{O}}}
\def\calm{{\cal{M}}}
\def\calmm{{\calm}^2}
\def\hl{h^0}
\def\ha{A^0}
\def\hh{H^0}
\def\hpm{H^\pm}
\def\mha{m_{\ha}}
\def\mhl{m_{\hl}}
\def\mhh{m_{\hh}}
\def\mhpm{m_{\hpm}}
\def\mz{m_Z}
\def\mw{m_W}
\def\mt{m_t}
\def\mb{m_b}
\def\msusy{M_{\rm SUSY}}
\def\msusyy{M_{\rm SUSY}^2}
\def\tr{\rm tr}
\def\cw{\cos\theta_W}
\def\sw{\sin\theta_W}
\def\swiv{\sin^4\theta_W}
\def\ctwob{\cos2\beta}
\def\ctwobb{\cos^2 2\beta}
\def\d1l{\delta\lambda_1}
\def\d2l{\delta\lambda_2}
\def\d3l{\delta\lambda_3}
\def\d4l{\delta\lambda_4}
\def\d5l{\delta\lambda_5}
\def\d6l{\delta\lambda_6}
\def\d7l{\delta\lambda_7}
\def\sb{\sin\beta}
\def\cb{\cos\beta}
\def\sbb{\sin^2\beta}
\def\cbb{\cos^2\beta}
\def\sbiv{\sin^4\beta}
\def\cbiv{\cos^4\beta}
\def\mzz{m_Z^2}
\def\mww{m_W^2}
\def\sw#1{\sin^{#1}\theta_W}
\def\cw#1{\cos^{#1}\theta_W}
%
%
% text
%
\section{Introduction}

With the LEP-2 Collider now beginning operation, a detailed
assessment of its discovery capabilities has recently been 
completed \cite{lep2workshop}.
%One of the objectives of this recent study is to recommend the
%ultimate center-of-mass energy for the machine.  The Higgs boson
%discovery reach is one of the key ingredients of this decision.
The search for the Higgs boson will play a central role in the LEP-2
program. 
A common rule of thumb states that for a center-of-mass
energy $\sqrt{s}$, a Higgs boson with $\mhl\lsim\sqrt{s}-100$~GeV can
be discovered (at the $5$-$\sigma$ level), assuming a sufficient integrated
luminosity.  (Slightly higher masses can be ruled
out at 95\% CL if no Higgs signal is seen.)  This means that with
$\sqrt{s}=192$~GeV, one can expect to rule out or discover the Higgs
boson if its mass is $\lsim\mz$.

In the Standard Model (SM), this Higgs mass discovery reach is not
particularly impressive.  Without any assumption about physics at
higher energy scales (above, say, 1 TeV), Higgs masses up to
about 700 GeV are possible (within the context of a ``weakly-coupled''
scalar sector).  On the other hand, if one assumes that no new
physics (beyond the SM) enters up to the Planck scale, then one
can deduce an upper bound of $\mhl\lsim 175$~GeV \cite{landau}.  In this light,
the LEP search is perhaps more significant.  But, in the SM, there
also exists a {\it lower} bound from stability arguments
\cite{quiros}.  Under
the assumption of no new physics up to the Planck scale, for
$\mt=175$~GeV, it follows that $\mhl\gsim 125$~GeV.  Moreover, a Higgs
boson discovery at LEP would imply that new physics must enter at
an energy scale below 100~TeV.  One of the leading candidates
for such new physics is low-energy supersymmetry.
\par
In the minimal supersymmetric extension of the Standard Model (MSSM),
the Higgs sector \cite{hhg}  consists of two Higgs doublets of
hypercharge $\pm
1$.  The scalar spectrum consists of two CP-even scalars, $\hl$ and
$\hh$ (with $\mhl\leq\mhh$), a CP-odd scalar $\ha$ and a charged
Higgs boson pair $\hpm$.
Unlike the SM, the Higgs self-couplings are not independent
parameters; but they are related to the electroweak gauge
couplings.  As a result, the Higgs mass spectrum is constrained.
At tree-level, all Higgs masses and couplings depend on two
parameters: $\mha$ and the ratio of Higgs vacuum expectation values,
$\tan\beta$.  The charged Higgs squared-mass is given by
$(\mhpm^2)_0=\mha^2+\mw^2$, while
the CP-even neutral Higgs squared-masses and corresponding mixing
angle $\alpha$ are obtained
by diagonalizing the $2\times 2$ mass matrix (in the hypercharge
basis)  
\begin{equation} \label{mssmtree}
\calm_0^2=
  \pmatrix{
   \mha^2\sin^2\beta+\mz^2\cos^2\beta & -(\mha^2+\mz^2)\sin\beta\cos\beta \cr
  -(\mha^2+\mz^2)\sin\beta\cos\beta & \mha^2\cos^2\beta+\mz^2\sin^2\beta\cr}\,,
\end{equation}
where the subscript 0 indicates tree-level quantities.
One can then prove that $(\mhl)_0\leq\mz|\cos2\beta|$.
If this tree-level inequality were reliable, then LEP-2 would have
the mass reach either to discover the Higgs boson or rule out the
MSSM!  When radiative corrections are taken into account,  the
Higgs mass bound increases \cite{hhprl,early-veff}.  
The dominant contribution to this
increase is a term of order $\Delta\mhl^2\sim 3g^2 m_t^4\log(M_{\tilde
t}^2/\mt^2)/(8\pi^2\mw^2)$, which arises due to an incomplete cancellation 
of top quark and top squark loops (the cancellation would be complete 
in the limit of exact supersymmetry).
\par
The radiative corrections to the Higgs mass have been computed by a
number of techniques, and using a variety of approximations such as
effective potential \cite{early-veff,veff,erz,carena} and 
diagrammatic methods \cite{hhprl,1-loop,madiaz,hempfhoang,completeoneloop}.
Complete one-loop diagrammatic computations of the MSSM Higgs masses
have been presented by a number of groups \cite{completeoneloop};
the resulting expressions are quite complex,
and depend on all the parameters of the MSSM.
Moreover, as noted above, the largest contribution to the one-loop radiative
corrections is enhanced by a factor of $m_t^4$ and grows
logarithmically with the top squark mass.  Thus, higher order
radiative corrections can be non-negligible for a large top
squark mass, in which case the large logarithms must be resummed.  The
renormalization group (RG) techniques for resumming the leading
logarithms has been developed by a number of 
authors \cite{rge,2loopquiros,llog}.
This procedure involves integrating a set of coupled partial
differential equations.  
As a result, the numerical evaluation of the one-loop Higgs masses
is time consuming and not very suitable for detailed phenomenological
analyses. 
The primary goal of this paper is to present a simple algorithm that
incorporates
the effects of the RG-improvement and minimizes the size of the
two-loop radiative corrections.  We then can apply this algorithm to
a suitable approximation to the full one-loop corrected Higgs masses.

We present a successive series of approximations to the one-loop
corrected Higgs masses, of increasing complexity, each one reflecting finer
details of the low-energy supersymmetric spectrum.  
Symbolically,
\begin{eqnarray}
\mhpm^2 & = & \left(\mhpm^2\right)_0+\left(\Delta \mhpm^2\right)_{\rm 1LL}
+\left(\Delta \mhpm^2\right)_{\rm mix}\,,\nonumber \\
\calm^2 & = & \calm_0^2+ \Delta\calm^2_{\rm 1LL}+ \Delta\calm^2_{\rm mix}\,,
\label{oneloopmasses}
\end{eqnarray}
where the subscript 0 refers to the tree-level result, the subscript
{\sl 1LL} refers to the
one-loop leading logarithmic approximation to the full one-loop
calculation, and the subscript {\sl mix} refers to the contributions
arising from $\widetilde q_L$--$\widetilde q_R$ mixing effects of
the third generation squarks.
The CP-even Higgs mass-squared eigenvalues are then given by
\begin{equation}
m^2_{\hh,\hl}=\frac{1}{2}\,\left[\calmm_{11}+\calmm_{22}\pm
\sqrt{[\calmm_{11}-\calmm_{22}]^2+4(\calmm_{12})^2}\,\right]\,,
\label{massev}
\end{equation}
and the corresponding mixing angle, $\alpha$, is
obtained from
\begin{eqnarray} \label{defalpha}
\sin 2\alpha & = & {2\calmm_{12}\over
\sqrt{[\calmm_{11}-\calmm_{22}]^2+4(\calmm_{12})^2}}\,,\nonumber \\
\cos 2\alpha & = & {\calmm_{11}-\calmm_{22}\over
\sqrt{[\calmm_{11}-\calmm_{22}]^2+4(\calmm_{12})^2}}\,.
\end{eqnarray}

In the first (and simplest) approximation, squark mixing effects are 
neglected and  the supersymmetric spectrum is characterized by one
scale, called $\msusy$.  We assume that $\msusy$ is sufficiently
large compared to $\mz$ such that logarithmically enhanced terms
at one-loop dominate over the non-logarithmic terms.\footnote{If this
condition does not hold, then the radiative corrections would
constitute only a minor perturbation on the tree-level predictions.}
In this case, the full
one-loop corrections ({\it e.g.}, obtained by a diagrammatic technique) 
are well approximated by the one-loop leading logarithmic approximation.
In this approximation, we neglect all squark mixing effects.
Explicit formulae for $\Delta\calmm_{\rm 1LL}$ and 
$(\Delta\mhpm^2)_{\rm 1LL}$ can be found in Appendix A.
The second approximation incorporates squark mixing effects.  These
are likely to be significant only in the third generation squark sector.
This approximation is parameterized by four supersymmetric parameters:
$\msusy$ (a common supersymmetric particle mass) and the third
generation squark mixing parameters: $A_t$, $A_b$ and $\mu$.  The
corresponding formulae for $\Delta\calmm_{\rm mix}$ and 
$(\Delta\mhpm^2)_{\rm mix}$ can be found in Appendix B.

The dominant contribution to the Higgs mass radiative corrections
enters through the exchange of the third generation squarks.
Thus, our third approximation treats this sector more precisely by
accounting for non-degenerate top and bottom squark masses.
This approximation is characterized by seven supersymmetric
parameters---the three squark mixing parameters mentioned above,
three soft-supersymmetry-breaking diagonal squark mass parameters,
$M_Q$, $M_U$, and $M_D$, and a common supersymmetry mass parameter
$\msusy$ which characterizes the masses of the first two generations
of squarks, the sleptons, the charginos, and the neutralinos.
A more precise set of formulae for  $\Delta\calmm_{\rm 1LL}$ and
$\Delta\calmm_{\rm mix}$
incorporating the detailed squark and slepton mass spectrum,
can be found in Appendix C.
Note that setting $M_Q=M_U=M_D=\msusy$ reduces this approximation to
the previous one.  
Finally, our fourth approximation incorporates
a non-trivial neutralino and chargino spectrum.  This introduces two
additional parameters, $M_1$ and $M_2$ which characterize the color singlet 
gaugino masses.  The higgsino masses are determined by $\mu$, which
already enters the analysis through the third generation squark mixing
effects.  The relevant correction terms to $\Delta\calmm_{\rm 1LL}$
incorporating a non-universal chargino and neutralino spectrum is
given in Appendix D.

\par

Given an approximation to the one-loop Higgs mass as described above
we then develop a simple algorithm to incorporate 
the leading effects of RG-improvement.  In Section 2, 
we demonstrate that the dominant higher order corrections can
be absorbed in the expressions for the one-loop corrected Higgs masses
by a suitable re-definition of $m_t$ (and $m_b$).  
%introduce a simple method for approximating the RG-improved
%masses.  At this point, the analysis is based on one-loop
%considerations. By incorporating the leading two-loop effects, we
%show how one can improve our approximation.  
Comparisons
between the results of our analytic approximation and the results of
the numerically integrated renormalization group equations (RGEs)
demonstrate the domain of
validity of our approximations.  In Section 3, we include the effects
of squark mixing.  We are able to modify our analytic approximation
in a simple way to incorporate the main squark mixing effects.
Section 4 summarizes the results of this paper.  In order to make
the paper self contained, we collect in a series of appendices 
the necessary formulae required to implement our algorithm.

This paper was motivated in part by the Higgs Bosons working group of 
the 1995 LEP-2 Workshop \cite{lep2higgs}, which examined in detail
the phenomenology of the light CP-even and CP-odd Higgs bosons (in the
MSSM).  The radiative corrections to the light CP-even Higgs mass play
a crucial role in determining the fraction of MSSM Higgs parameter space
accessible to LEP-2 at its maximum energy of $\sqrt{s}\simeq 192$~GeV.
As a result, the graphs we present in this paper focus on the light
CP-even Higgs mass, although our formulae can also be used to compute
the radiatively corrected masses of the heavy CP-even and charged Higgs
bosons.  While we were completing this work, Carena and collaborators
published two papers in which analytic approximations to the radiatively
corrected MSSM Higgs masses are also developed \cite{carena}.
Their methods and emphasis are somewhat different from ours.   
Nevertheless, the final results are quite similar, and our numerical
work (in cases where we have compared) typically agree to within
1~GeV in the evaluation of Higgs masses.  
Some of the results of our work have been previously reported in 
Refs.~\cite{lep2higgs} and \cite{brussels}, 
and our formulae have been employed in the review of Gunion, Stange, and 
Willenbrock in Ref.~\cite{gunionrev}. 
\par

\section{RG-improved Higgs Masses---No Squark Mixing}

In this section, we examine the simplest case in which the
supersymmetric spectrum is characterized by a single scale called
$\msusy$.  We begin with the one-loop leading logarithmic expressions
for the Higgs squared-masses given in Appendix A.  
Note that for $\msusy\gg m_Z$, the logarithmically
enhanced terms appearing in the formulae of Appendix A
can potentially spoil the perturbative expansion.
In this case, it is necessary to perform a RG-improvement which
resums the leading logs to all orders in perturbation theory.  The
resulting RG-improved perturbative expansion is better behaved and
more reliable.  The numerical effects of the RG-improvement can be
significant for values of $\msusy$ as low as 500~GeV.
%In this paper, RG-improved masses will be denoted by an overline
%({\it e.g.}, $\overline{\calmm}$ denotes the RG-improved CP-even
%neutral Higgs squared-mass matrix). 

%\par
%\begin{figure}
%\vspace*{13pt}
%%\leftline{\hfill\vbox{\hrule width 5cm height0.001pt}\hfill}
%\vspace*{8cm}             %ORIGINAL SIZE=1.6TRUEIN x 100% - 0.2TRUEIN
%\special{psfile=hmass3.ps
%  voffset= -40 hoffset= 445 hscale=55 vscale=47 angle = 90}
%%\leftline{\hfill\vbox{\hrule width 5cm height0.001pt}\hfill}
%\caption{
%This is the 1-plot try.}
%\label{fig2}
%\end{figure}

Numerical integration of the coupled RGEs is a straightforward but
time-consuming process.   Here, we develop a simple analytic formula
that closely reproduces the result of the full numerical computation
over the parameter regime of interest.  Specifically, we are
interested in values of $\msusy$ that lie between 200~GeV and 2~TeV.
For values of $\msusy$ approaching $m_Z$, the 
leading logarithmic corrections are of the same size as non-leading
corrections not included in the Appendix A formulae; however both
are small corrections to the tree-level predictions.  Moreover,
in this regime, the effects of the RG-improvement
are insignificant and can be neglected.  For values of $\msusy$
above 2~TeV, the supersymmetry breaking scale is becoming unnaturally
large (compared to the scale of electroweak symmetry breaking).
In this section, we shall apply our technique to the one-loop leading
logarithmic CP-even Higgs squared-mass matrix, $\calmm_{\rm 1LL}\equiv
\calmm_0+\Delta\calmm_{\rm 1LL}$.  
%In Section 4, we
%will indicate how to apply our method to the full one-loop corrected
%squared-mass matrix $\calmm_{\rm 1LC}$.
\par
The matrix $\calmm_{\rm 1LL}$ depends explicitly on the top quark
mass.  But, which top-quark mass should one use?  In a diagrammatic
analysis, working in an on-shell scheme, one would use the pole mass.
The analysis based on RG-running would naturally use the running
mass, $m_t(m_t)$.
The choice between the pole mass and $m_t(m_t)$ cannot be decided
based on one-loop considerations alone.
Since the dependence on $m_t$ enters only at one-loop, the
distinction between various definitions of $m_t$ is a two-loop
effect.  We will return to this distinction later in this
section.  Dependence on $m_b$ will be considered below as well, although the
numerical distinction among different $m_b$ choices is small. 
\par
We can now state a simple analytic formula that incorporates the
dominant effects of the RG-improvement:
\begin{equation}
\calmm_{\rm 1RG}\simeq\overline{\calmm}_{\rm 1LL}\equiv
\calmm_{\rm 1LL}\left(m_t(\mu_t),m_b(\mu_b)\right)\,,\qquad 
\mu_t\equiv\sqrt{\mt\msusy}\,,\qquad\mu_b\equiv\sqrt{\mz\msusy}\,.
\label{simpleform}
\end{equation}
That is, we assert that the numerically integrated 
RG-improved CP-even Higgs
squared-mass matrix, $\calmm_{\rm 1RG}$, 
is well approximated by replacing all
occurrences of $m_t$ and $m_b$ in $\calmm_{\rm 1LL}(m_t,m_b)$ by
the corresponding running masses.\footnote{In this section, an overline
above a quantity will indicate that the replacement of $m_t$ [$m_b$]
by $m_t(\mu_t)$ [$m_b(\mu_b)$] has been made.}
Before justifying this assertion, we need formulae for $m_b(\mu)$ and
$m_t(\mu)$.  First, consider $\mha={\cal O}(\mz)$.  In this case,
at mass scales below $\msusy$, the effective theory of the Higgs
sector is that of a non-supersymmetric two-Higgs-doublet model (2HDM).
In this model, the
quark mass is the product of the Higgs-quark Yukawa coupling
($h_q$) and the appropriate Higgs vacuum expectation value:
\begin{eqnarray}
m_b(\mu) & = & \frac{1}{\sqrt{2}}\,h_b(\mu)\,v_1(\mu)\,,\nonumber \\
m_t(\mu) & = & \frac{1}{\sqrt{2}}\,h_t(\mu)\,v_2(\mu)\,,
\label{topmass}
\end{eqnarray}
where we employ the normalization $v_1^2+v_2^2=4m_W^2/g^2=(246~{\rm GeV})^2$.
At scales $\mu\leq\msusy$, we employ the one-loop non-supersymmetric
RGEs of the 2HDM (see {\it e.g.}, Ref.~\cite{yukrges}) for $h_b$,
$h_t$ and the vacuum expectation values.
%\begin{eqnarray}
%&&\beta_{h_b^2}\equiv\frac{\rm d}{\rm d\ln\mu^2}\,h_b^2
% = \frac{1}{16\,\pi^2}\left[\,\frac{9}{2} h_b^2
%  +\frac{1}{2} h_t^2-8 g_s^2-\frac{9}{4} g^2-
%  \frac{5}{12} g^{\prime 2}\,\right]\,h_b^2\,,
%\nonumber \\
%&&\beta_{h_t^2}\equiv\frac{\rm d}{\rm d\ln\mu^2}\,h_t^2
% = \frac{1}{16\,\pi^2}\left[\,\frac{9}{2} h_t^2
%  +\frac{1}{2} h_b^2-8 g_s^2-\frac{9}{4} g^2-
%  \frac{17}{12} g^{\prime 2}\,\right]\,h_t^2\,,
%\nonumber \\
%&&\beta_{v_1^2}\equiv\frac{\rm d}{\rm d\ln\mu^2}\,v_1^2  
% = \frac{1}{64\,\pi^2}\left[\,9 g^2+3 g^{\prime 2}
%%  -12 h_b^2\,\right]\,v_1^2\,, \nonumber \\
%&&\beta_{v_2^2}\equiv\frac{\rm d}{\rm d\ln\mu^2}\,v_2^2  
% = \frac{1}{64\,\pi^2}\left[\,9 g^2+3 g^{\prime 2}
%  -12 h_t^2\,\right]\,v_2^2\,,
%\label{rges}
%\end{eqnarray}
This yields
\begin{eqnarray}
&&\frac{\rm d}{\rm d\ln\mu^2}\,m_b^2 =
\frac{1}{64\,\pi^2}\,\left[\,6 h_b^2+2 h_t^2-32 g_s^2
+\frac{4}{3} g^{\prime 2}\,\right]\,m_b^2\,, \nonumber \\
&&\frac{\rm d}{\rm d\ln\mu^2}\,m_t^2 =
\frac{1}{64\,\pi^2}\,\left[\,6 h_t^2+2 h_b^2-32 g_s^2
-\frac{8}{3} g^{\prime 2}\,\right]\,m_t^2\,.
\label{mtrge}
\end{eqnarray}
For $\mha={\cal O}(\msusy)$, the effective theory of the Higgs
sector at mass scales below $\msusy$ 
is that of the one-Higgs doublet Standard Model.  In this case,
we define $m_q(\mu)=h_q^{\rm SM}(\mu) v(\mu)/\sqrt{2}$, where 
$v(\mz)\simeq 246$~GeV is the
one-Higgs-doublet Standard Model vacuum expectation value.  In this
case eq.~(\ref{mtrge}) is modified by replacing $6h_t^2+2h_b^2$
with $6(h_t^{\rm SM})^2-6(h_b^{\rm SM})^2$ in the RGE for $m_t^2$
(and interchange $b$ and $t$ to obtain the RGE for $m_b^2$).

To solve these equations, we also need the evolution equations of
$g_s$, and $g^\prime$.  But, an approximate solution is
sufficient for our purposes.  Since $g^\prime$ is small, we drop it.
We do not neglect the $h_b$ dependence
which may be significant if $\tan\beta$ is large.
Then, we can iteratively solve eq.~(\ref{mtrge}) to one loop by 
ignoring the $\mu$ dependence of the right hand side.  We find
\begin{equation}
m_t(\mu) = m_t(m_t)\times
\cases{1-{1\over\pi}\left(\alpha_s-{1\over 16}
(\alpha_b+3\alpha_t)\right)\,
\ln\left(\mu^2/m_t^2\right)\,, & $\mha\simeq{\cal
O}(\mz)\,,$\cr
1-{1\over\pi}\left(\alpha_s-{3\over 16}
(\alpha_t^{\rm SM}-\alpha_b^{\rm SM})\right)\,
\ln\left(\mu^2/m_t^2\right)\,, & $\mha\simeq{\cal
O}(\msusy)\,,$\cr}
\label{mtrun}
\end{equation}
where $\alpha_t\equiv h_t^2/4\pi$, {\it etc.}, and
all coupling on the right hand side are evaluated at $m_t$.
Similarly,
\begin{equation}
m_b(\mu) = m_b(\mz)\times
\cases{1-{1\over\pi}\left(\alpha_s-{1\over 16}
(\alpha_t+3\alpha_b)\right)\,
\ln\left(\mu^2/\mz^2\right)\,,&$\mha\simeq{\cal
O}(\mz)\,,$ \cr
1-{1\over\pi}\left(\alpha_s-{3\over 16}
(\alpha_b^{\rm SM}-\alpha_t^{\rm SM})\right)\,
\ln\left(\mu^2/\mz^2\right)\,,&$\mha\simeq{\cal
O}(\msusy)\,,$\cr}
\label{mbrun}
\end{equation}
For intermediate values of $\mha$, one may
extrapolate the above formulae between the two regions indicated
in the step function approximation.
Using the results of eqs.~(\ref{mtrun}) and (\ref{mbrun}) in
eq.~(\ref{simpleform}), and 
diagonalizing the squared-mass matrix of eq.~(\ref{simpleform})
yields our approximation to the RG-improved one-loop
neutral CP-even Higgs squared-masses. 
 
Before justifying the above results, we exhibit a numerical comparison
among various computations of the one-loop corrected 
light CP-even Higgs mass.
First, we evaluate two expressions for the RG-unimproved
one-loop Higgs mass---the one-loop leading log Higgs mass calculated
from $\calmm_{\rm 1LL}$ and from a simplified version of $\calmm_{\rm 1LL}$ 
in which only the dominant
terms proportional to $m_t^4$ are kept.  In the latter
case, we denote the CP-even Higgs squared-mass matrix by
$\calmm_{\rm 1LT}\equiv \calmm_0+\Delta\calmm_{\rm 1LT}$, where
\begin{equation}
\Delta\calmm_{\rm 1LT}\equiv {3g^2 m_t^4\over 8\pi^2 m_W^2 \sbb}
 \ln\left(\msusyy/\mt^2\right)
\left(\begin{array}{cc}
  0 & 0 \\
  0 & 1
\end{array}\right)\,.
\label{topapprox}
\end{equation}

In many analyses of $\Delta\calmm_{\rm 1LT}$ and $\Delta\calmm_{\rm 1LL}$ 
that have appeared previously
in the literature, the Higgs mass radiative corrections were evaluated with 
the pole mass, $m_t$.   Some have argued that one should take $m_t$ to be
the running mass evaluated at $m_t$, although to one-loop accuracy, the two
choices cannot be distinguished.  Nevertheless, because the leading
radiative effect is proportional to $m_t^4$, the choice of $m_t$ in the
one-loop formulae is numerically significant, and can lead to differences
as large as 10 GeV in the computed Higgs mass.  
At the end of this section, we justify 
the choice of using $m_t(m_t)$ as opposed to $m_t^{\rm pole}$
(prior to RG-improvement) 
by invoking information from a two-loop analysis.  
Thus, our numerical results for the light CP-even Higgs mass before
RG-improvement is significantly lower (when $\msusy$ is large)
as compared to the original computations given in the literature, for fixed
$m_t^{\rm pole}$.  In this paper, we have taken $m_t(m_t)=166.5$~GeV in all
of our numerical work.
We then apply our algorithm for RG-improvement by replacing $m_t$ and
$m_b$ by their running masses evaluated at 
$\mu_t$ and $\mu_b$, respectively, as
specified in eq.~(\ref{simpleform}).  

We now show examples for $\mha=1$~TeV and two choices of $\tanb$
in Figs.~\ref{hhhfig1} and \ref{hhhfig2}, and for $\mha=100$~GeV and
$\tanb=20$ in Fig.~\ref{hhhfig3}.\footnote{For $\mha=100$~GeV and $\tanb=1.5$,
the resulting light Higgs mass lies below experimental Higgs mass bounds
obtained by the LEP collaborations \cite{higgslimits}.}
Each plot displays five predictions for $\mhl$ based on
the following methods for computing the Higgs squared-mass matrix:
(i) $\calmm_{\rm 1LT}$; (ii) $\calmm_{\rm 1LL}$; 
(iii) $\overline{\calmm}_{\rm 1LT}$ (iv) $\overline{\calmm}_{\rm 1LL}$;
and (v) $\calmm_{\rm 1RG}$ [see footnote below eq.~(\ref{simpleform})].
The following general features are
noteworthy.  First, we observe that over the region of $\msusy$ shown,
$\calmm_{\rm 1RG}\simeq\overline{\calmm}_{\rm 1LL}$.  In fact, $\mhl$
computed from $\overline{\calmm}_{\rm 1LL}$ is within 1 GeV of the
numerical RG-improved $\mhl$ in all sensible
regions of the parameter space
($1\leq\tan\beta\leq m_t/m_b$ and $m_t$, $\mha\leq\msusy
\leq 2$~TeV).  For values of $\msusy>2$~TeV, the Higgs masses obtained
from $\overline{\calmm}_{\rm 1LL}$ begin
to deviate from the numerically integrated RG-improved result.
Second, the difference between $\mhl$ computed from $\calmm_{\rm 1LL}$
and from $\calmm_{1RG}$ is non-negligible for large values of $\msusy$;
neglecting RG-improvement can lead to an overestimate of $\mhl$ which
in some areas of parameter space can be as much as 10 GeV. Finally,
note that although the simplest approximation of $\mhl$ based on
$\calmm_{\rm 1LT}$ reflects the dominant radiative corrections, it yields
the largest overestimate of the light Higgs boson mass.

\begin{figure}[htb]
%\centerline{ \psfig{file=hhhfig1.ps,width=8.3cm,height=6.8cm,angle=90}
%\hfill      \psfig{file=hhhfig2.ps,width=8.5cm,height=6.8cm,angle=90}}
\centerline{\psfig{file=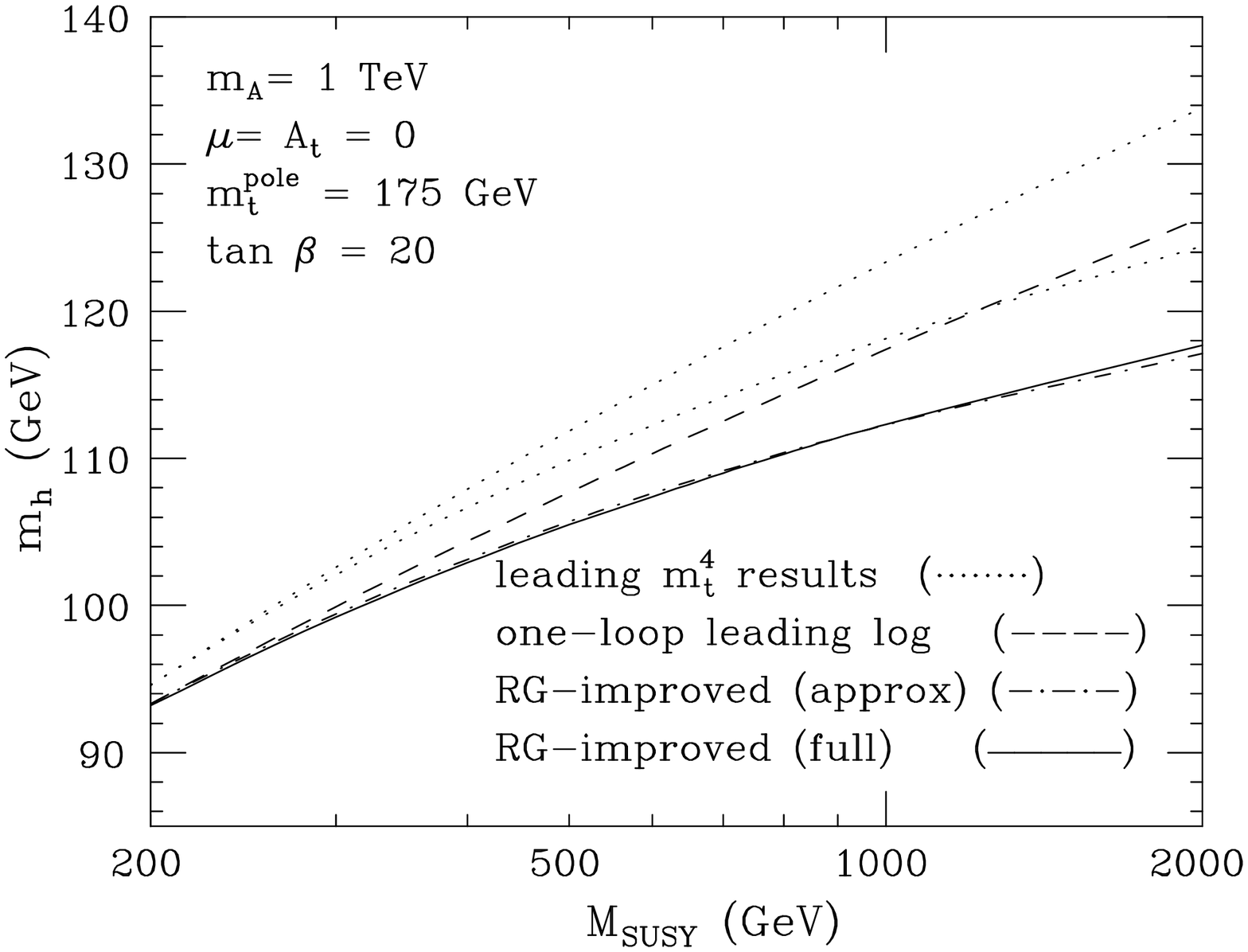,width=12cm,height=9.5cm,angle=90}}
\fcaption{The radiatively corrected light CP-even Higgs mass is plotted
as a function of $\msusy$ for $\tan\beta=20$ and $\mha= 1$~TeV.
The one-loop leading logarithmic computation [dashed line] 
is compared with the RG-improved result which was obtained
by numerical analysis [solid line] and by using the simple analytic 
result given in eq.~(\protect\ref{simpleform}) [dot-dashed line].  
For comparison, the
results obtained using the leading $m_t^4$ approximation of
eq.~(\protect\ref{topapprox}) [higher dotted line], and its RG-improvement
[lower dotted line] are also exhibited.  $\msusy$ characterizes the
scale of supersymmetry breaking and can be regarded (approximately)
as a common supersymmetric scalar mass; squark mixing effects are
set to zero.  The running top quark mass used in
our numerical computations is $m_t(m_t)= 166.5$~GeV.}
\label{hhhfig1}
\end{figure}

\begin{figure}[hp]
%\centerline{ \psfig{file=hhhfig1.ps,width=8.3cm,height=6.8cm,angle=90}
%\hfill      \psfig{file=hhhfig2.ps,width=8.5cm,height=6.8cm,angle=90}}
\centerline{\psfig{file=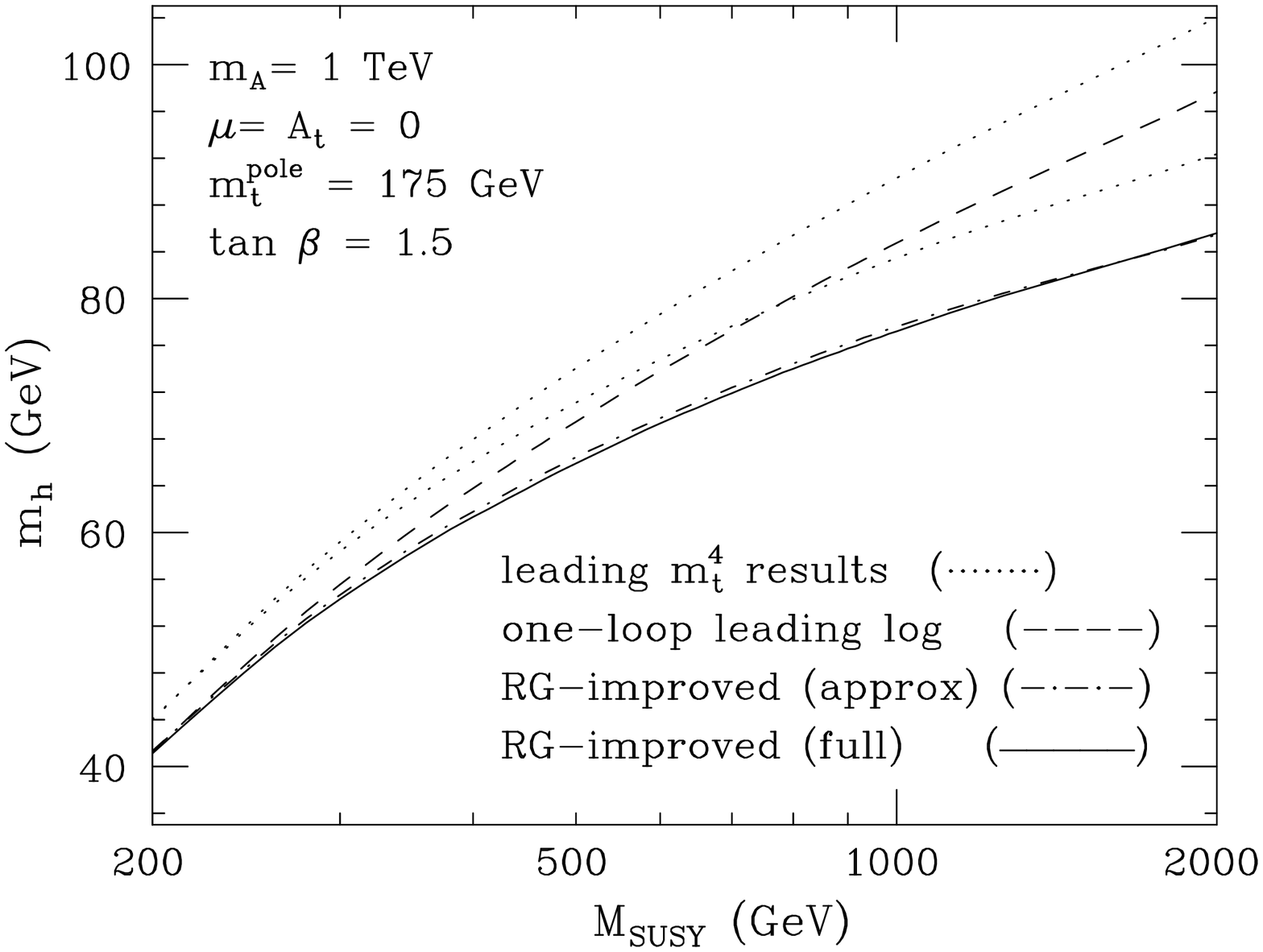,width=12cm,height=9.5cm,angle=90}}
\fcaption{The radiatively corrected light CP-even Higgs mass is plotted
as a function of $\msusy$ for $\tan\beta=1.5$ and $\mha= 1$~TeV.
See the caption to Fig.~\protect\ref{hhhfig1}.}
\label{hhhfig2}
%\end{figure}
\vspace*{1pc}
%
%\begin{figure}[htb]
%\centerline{ \psfig{file=hhhfig1.ps,width=8.3cm,height=6.8cm,angle=90}
%\hfill      \psfig{file=hhhfig2.ps,width=8.5cm,height=6.8cm,angle=90}}
\centerline{\psfig{file=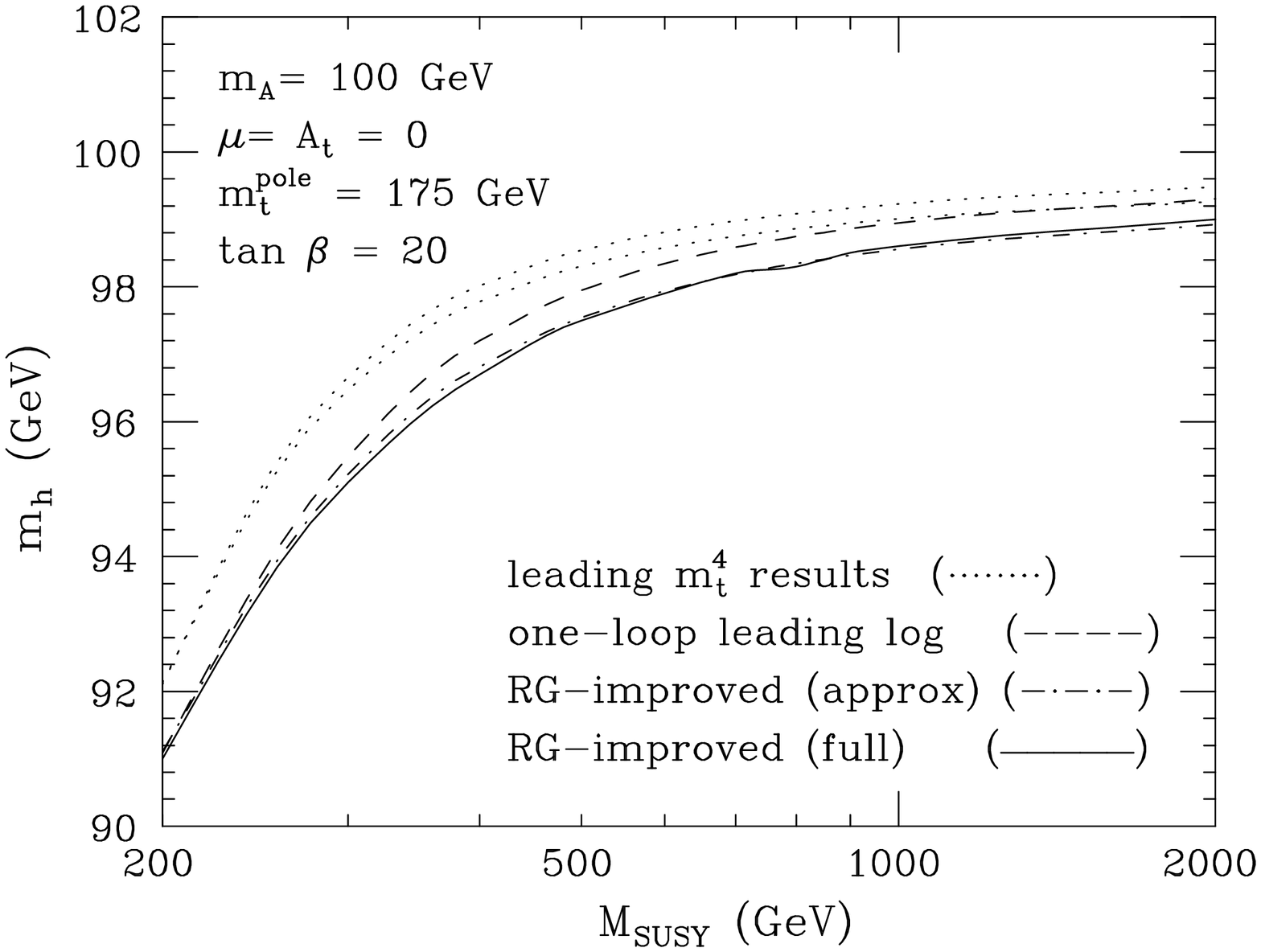,width=12cm,height=9.5cm,angle=90}}
\fcaption{The radiatively corrected light CP-even Higgs mass is plotted
as a function of $\msusy$ for $\tan\beta=20$ and $\mha= 100$~GeV.
See the caption to Fig.~\protect\ref{hhhfig1}.}
\label{hhhfig3}
\end{figure}

\par
We now give a justification for the prescription proposed in 
eq.~(\ref{simpleform}) for the radiatively corrected CP-even Higgs
squared-mass matrix.
Because the one-loop leading logarithmic correction proportional to
$m_t^4$ is dominant, it is
sufficient to consider the two-loop leading and next-to-leading
log contributions proportional to
$m_t^4\,\alpha_s$ and $m_t^4\,\alpha_t$. To simplify the presentation we will
exhibit the calculational steps for the case $\mha\sim\calo(\msusy)$,
$m_b=0$, and $\beta=\pi/2$.
%\footnote{Because the
%$m_t^4\alpha_t$ and $m_t^4\,\alpha_s$ contributions occur only in the
%$\calmm_{22}$ element of the neutral CP-even Higgs squared-mass matrix,
%our conclusions remain valid for arbitrary values of $\mha$ and $\tan\beta$.}
%To keep the formulae as simple as possible we do not indicate
%explicitly terms beyond the two loop level.
%\par
In this case, the effective Higgs sector at scales below $\msusy$ is
the one-Higgs-doublet Standard Model, which is characterized by one
Higgs self-coupling parameter $\lambda$.  Supersymmetry fixes the
value of $\lambda$ above the supersymmetry breaking scale, $\msusy$,
to be  
\begin{equation}
\lambda(\msusy) = \frac{1}{4}(g^2+g^{\prime 2})(\msusy)\,.
\label{lambdaatmsusy}
\end{equation}
The physical Higgs mass is determined at the scale $\mz$ (assumed to be
much less than $\msusy$): $\mhl=\lambda(\mz)v^2(\mz)$.  Note that if
supersymmetry were unbroken, then $\lambda$ would be fixed to the
value given in eq.~(\ref{lambdaatmsusy}), which would imply that
$\mhl=\mz$.  This is the expected result for $\beta=\pi/2$ and
$\mha>\mz$ [see eq.~(\ref{mssmtree})].  In the case of broken
supersymmetry, eq.~(\ref{lambdaatmsusy}) is taken as the boundary
condition for the one-Higgs doublet Standard Model evolution of
$\lambda$ from $\msusy$ down to $\mz$.  
We work in the approximation $h_b=g=g^\prime=0$ 
(a more precise computation is not required here).  
In this approximation, we employ the following one-loop $\beta$-functions
\begin{eqnarray}
&&\beta_{\lambda}\equiv\frac{\rm d}{\rm d\ln\mu^2}\,\lambda
 = \frac{3}{8\pi^2}\left[\,\lambda^2-h_t^4\right]-2\lambda\gamma_v\,,
\nonumber \\
&&\beta_{h_t^2}\equiv\frac{\rm d}{\rm d\ln\mu^2}\,h_t^2
 = \frac{1}{16\pi^2}\left[\,\frac{9}{2} h_t^2 -8 g_s^2\right]\,h_t^2\,,
\label{rges}
\end{eqnarray}
where
\begin{equation} \label{anomdim}
\gamma_v\equiv{1\over v^2}\frac{\rm d}{\rm d\ln\mu^2}\,v^2=  
 \frac{-3}{16\pi^2}h_t^2\,.
\end{equation}
Solving iteratively the RGE for $\lambda$, and
inserting the result in $\mhl^2=\lambda(\mz)v^2(\mz)$, we end up 
with\footnote{Note that in
the approximation where $h_b=g=g'=0$ and $\lambda$ is neglected compared
with $h_t$, both $\lambda(\mu)$ and $v(\mu)$ do not run for $\mu\leq\mt$.}
\begin{equation}
\mhl^2= \mz^2+{3g^2 m_t^4(m_t)\over 8\pi^2\mw^2}\ln\left(
{\msusyy\over\mt^2}\right)\left[1+\left(\gamma_v+
{\beta_{h_t^2}\over h_t^2}\right)\ln\left({\msusyy\over\mt^2}\right)
\right]\,,
\label{higgsmass1}
\end{equation}
where $h_t\equiv h_t(m_t)$ and
$m_t(m_t)=h_t\,v(m_t)/\sqrt{2}$.  Clearly, the term in
brackets can be absorbed into the factor of $m_t^4$, thereby
converting $m_t(m_t)$ into the running top quark mass evaluated at a
new scale $\mu_t$.  Comparison
with eq.~(\ref{mtrun}) shows that eq.~(\ref{higgsmass1}) can be
rewritten as
\begin{equation}
\mhl^2= \mz^2+{3g^2 m_t^4(\mu_t)\over 8\pi^2\mw^2}\ln\left(
{\msusyy\over\mt^2}\right)\,,
\label{higgsmass2}
\end{equation}
where $\mu_t\equiv\sqrt{\mt\msusy}$.  
Note that in the one-loop leading log analysis presented above, there is no
distinction between $m_t(m_t)$ [$\mhl(\mz)$] and the top quark 
[Higgs boson] pole mass.  Moreover, it is unclear at what scale to
evaluate the factor of $m_t$ appearing in the argument of the logarithm in
eq.~(\ref{higgsmass2}).  These ambiguities can be resolved by a
two-loop analysis.

Ref.~\cite{hempfhoang} used a diagrammatic technique to
compute the leading two-loop contributions to the Higgs pole mass in
terms of the top quark pole mass.  In the approximation that
$h_b=g=g'=0$ and $\msusy\gg\mt$ (and in the absence of squark mixing), 
the result of this computation can
be written in the following form\footnote{In our two-loop analysis, we
have defined $g^2/m_W^2\equiv 4\sqrt{2}G_\mu$, where
$G_\mu=1.16639\times 10^{-5}~{\rm GeV}^{-2}$ is the
muon decay constant \cite{pdg} and $\mw$ is the physical
$W^\pm$ mass.}
\begin{equation} \label{leading2loop}
\mhl^2 = \mz^2+
  \frac{3g^2}{8\pi^2\mw^2}\,
  (m_t^{\rm pole})^4\,
    \ln\left(\frac{\msusyy}{m_t^2}\right)\,
  \left[1+\left(\gamma_v+
        \frac{\beta_{h_t^2}}{h_t^2}\right)\,
        \ln\left(\frac{\msusyy}{m_t^2}\right)-
        \left(\frac{4\alpha_s}{\pi}\right)+
        \frac{13}{8}\left(\frac{\alpha_t}{\pi}\right)\,\right] \,.
\end{equation}
In the same approximation as above,
the top quark pole mass can be expressed (at one loop)
in terms of the running top quark mass evaluated at 
$m_t$ \cite{tpoleref1,tpoleref2} 
\begin{equation} \label{tpole}
m_t^{\rm pole}  =  m_t(m_t)\,\left[1+
    \frac{4}{3}\left(\frac{\alpha_s}{\pi}\right)-
    \frac{1}{2}\left(\frac{\alpha_t}{\pi}\right)\right]\,,
\end{equation}
where $m_t(m_t)$ is defined below eq.~(\ref{higgsmass1}).\footnote{We 
caution the reader that Ref.~\cite{tpoleref2} defines
$m_t(m_t)=2^{-3/4}G_\mu^{-1/2}h_t(m_t)$, which differs slightly from 
the definition of $m_t(m_t)$ used in our two-loop analysis.}
Inserting the above result into eq.~(\ref{leading2loop}) yields
\begin{equation} \label{twolooprun} 
\mhl^2 = \mz^2+
  \frac{3g^2}{8\pi^2\mw^2}\,
  m_t^4(m_t)\,
    \ln\left(\frac{\msusyy}{m_t^2}\right)\,
  \left[1+\left(\gamma_v+
        \frac{\beta_{h_t^2}}{h_t^2}\right)\,
        \ln\left(\frac{\msusyy}{m_t^2}\right)+
        \frac{4}{3}\left(\frac{\alpha_s}{\pi}\right)-
        \frac{3}{8}\left(\frac{\alpha_t}{\pi}\right)\,\right] \,.
\end{equation}
Given the Higgs mass computed to two-loop next-to-leading logarithmic
accuracy, consider the error one makes by using
eq.~(\ref{higgsmass1}), where the two-loop
next-to-leading logarithms are neglected.  
The size of the error depends on whether 
one uses the top quark pole mass or $m_t(m_t)$.  By comparing 
eqs.~(\ref{leading2loop}) and (\ref{twolooprun}), we see that 
the two-loop next-to-leading log term is a $10\%$ correction if 
the top quark pole mass is used, whereas it is a $3\%$ correction
if $m_t(m_t)$ is used.\footnote{In Ref.~\cite{hempfhoang} it is
shown that the two-loop non-logarithmic piece is negligible.}
This is our justification for using the running
top-quark mass in our one-loop analysis.

One can also derive eq.~(\ref{twolooprun}) using the RG-techniques
employed above.  Only three modifications of our one-loop analysis
are required.  First, we must
distinguish between the Higgs pole mass (denoted henceforth by $\mhl$ with
no argument) and the running Higgs mass evaluated at $\mhl$.
%$\mhl^2=\lambda(\mhl)v^2(\mhl)$.  
Using the results of Sirlin and Zucchini \cite{sirlin}
(in the limit of $h_b=g=g'=0$ and $\lambda\ll h_t$), 
\begin{equation}
\mhl^2  =  {4\,\mw^2\,\lambda(\mhl)\over g^2}\left[1+\frac{1}{8}
 \left(\frac{\alpha_t}{\pi}\right)\right]\,.
\label{poletorun}
\end{equation}
Second, we need only the $h_t$ and $g_s$ dependent parts of the 
two loop contribution to
$\beta_\lambda$.  That is, $\beta_\lambda$ of eq.~(\ref{rges}) is
modified as follows\footnote{The numerical impact of the two-loop
RGEs for the Higgs mass computation was examined in Ref.~\cite{2loopquiros} 
and shown to be a few percent of the one-loop radiative 
corrections.} \cite{2looprges}
\begin{equation}
\beta_{\lambda} \longrightarrow \beta_{\lambda} +
  \frac{1}{(16\pi^2)^2}\,\left[\,30 h_t^6-32 h_t^4\,g_s^2\,\right]\,.
\label{betalambdatwoloop}
\end{equation}
Third, we must take into account a finite correction
between $v^2(\mt)$ and $v^2\equiv 4\mw^2/g^2$,
\begin{equation} \label{vevcorrection}
v^2(\mt)={4\mw^2\over g^2}\left[1-{3\over 8}\left({\alpha_t\over\pi}\right) 
\right]\,.
\end{equation}
This result can be deduced, for example, from Ref.~\cite{hempfhoang}. 
Repeating the procedure leading to eq.~(\ref{higgsmass1}) using the 
modifications indicated in eqs.~(\ref{poletorun})--(\ref{vevcorrection}),
we easily reproduce eq.~(\ref{twolooprun}).  

Finally, we can address the question of which
top quark mass should appear in the argument of the logarithm.  Using
eq.~(\ref{mtrun}) to convert
the $m_t^2$ in the argument of the logarithm to $m_t^2(\mu_t)$, we find
\begin{equation} 
\mhl^2  =  \mz^2+
  \frac{3g^2}{8\pi^2\mw^2}\,
  m_t^4(\mu_t)\, \ln\left(\frac{\msusyy}{m_t^2(\mu_t)}\right)\,
  \left[1+
        \frac{1}{3}\left(\frac{\alpha_s}{\pi}\right)-
        \frac{3}{16}\left(\frac{\alpha_t}{\pi}\right)\,\right]\,.
\label{higgsmass3}
\end{equation}
One can check that the sum of the terms in the brackets
deviates from one by less than $1\%$. 

Since the leading $m_t^4$ term
provides the dominant source of the neutral Higgs mass radiative
correction, it follows that our algorithm of replacing $m_t$ with
$m_t(\mu_t)$ in the one-loop formula for $\calmm_{\rm 1LL}$
successfully reproduces the most important aspects of the
RG-improvement while minimizing the effects of the
non-leading-logarithmic two-loop effects, which are numerically small.
Note that in the numerical work, we use eq.~(\ref{mtrun}) to compute
the running top-quark mass at the scale of $\mu_t$ in terms of 
$m_t(m_t)$.  The latter quantity is expressed in terms of the top
quark pole mass using the two-loop QCD corrections of Ref.~\cite{tpoleref1}
(approximately a $6\%$ effect) and the electroweak corrections of
Ref.~\cite{tpoleref2} (a $-0.4\%$ correction for 
$m_t=175$~GeV and $\mhl\sim\mz$).\footnote{Since we are computing the
radiatively corrected Higgs mass for arbitrary $\mha$ and $\tanb$, it is
no longer appropriate to simply use eq.~(\ref{tpole}) for 
the top quark pole mass in terms of $m_t(m_t)$.}
For $m_t^{\rm pole}=175$~GeV, this yields roughly $m_t(m_t)= 166.5$~GeV.

We may also apply our algorithm to the radiatively corrected charged
Higgs mass.  One can check that the twice-iterated two-loop leading
log proportional to $m_t^2m_b^2$ is correctly reproduced.  However, in
contrast to the one-loop radiatively corrected neutral Higgs mass,
there are no one-loop leading logarithmic corrections to $\mhpm^2$ that
are proportional to $m_t^4$.  Thus, we expect that our charged Higgs
mass approximation will not be quite as reliable as our neutral Higgs mass 
approximation.

\section{RG-Improved Higgs Masses---Squark Mixing Effects Included}

In the previous section, we focused on the radiative corrections
under the assumption that one mass scale, $\msusy$ characterizes the
supersymmetric masses.  This is probably not a very realistic
assumption.  
%A detailed treatment of thresholds in the RG-approach
%was given in ref. \cite{llog}.  
In this section, we focus on the
effects arising from the mass splittings and $\widetilde
q_L$--$\widetilde q_R$ mixing
in the third generation
squark sector.  The latter can generate additional squared-mass shifts
proportional to $m_t^4$ and thus can 
have a profound impact on the radiatively
corrected Higgs masses.\footnote{The effects of mass splittings from
other supersymmetric sectors to the radiatively corrected Higgs masses
are quite small and will be neglected in
the numerical analysis of this paper.  However, in order to be complete,
we present in Appendix D analytic expressions for the contributions from 
mass splittings in the chargino and neutralino sectors.} 
In order to perform a RG-improvement of these effects, we
follow the approach of Ref.~\cite{llog}.
First, we define our notation (we follow the
conventions of Ref.~\cite{pdg2}).  In third family notation, the squark
mass eigenstates are obtained by diagonalizing the following two
$2\times 2$ matrices.  The top-squark squared-masses are eigenvalues of
\begin{equation}
\left(\begin{array}{cc}
  M_{Q}^2+m_t^2+t_L m_Z^2 & m_t X_t \\
  m_t X_t & M_{U}^2+m_t^2+t_R m_Z^2
\end{array}\right)  \,,
\label{stopmatrix}
\end{equation}
where $X_t \equiv A_t-\mu\cot\beta$,
$t_L\equiv ({1\over 2}-e_t\sin^2\theta_W)\cos2\beta$ and
$t_R\equiv e_t\sin^2\theta_W\cos2\beta$, with $e_t=+2/3$.
The bottom-squark squared-masses are eigenvalues of
\begin{equation}
\left(\begin{array}{cc}
  M_{Q}^2+m_t^2+b_L m_Z^2 & m_b X_b \\
 m_b X_b & M_{D}^2+m_t^2+b_R m_Z^2
\end{array}\right) \,,
\label{sbotmatrix}
\end{equation}
where $X_b \equiv A_b -\mu\tan\beta$,
$b_L\equiv (-{1\over 2}-e_b\sin^2\theta_W)\cos2\beta$ and
$b_R\equiv e_b\sin^2\theta_W\cos2\beta$, with $e_b=-1/3$.
$M_{Q}$, $M_{U}$, $M_{D}$, $A_t$, and
$A_b$ are soft-supersymmetry-breaking parameters,
and $\mu$ is the supersymmetric Higgs mass parameter. 
We treat the squark mixing perturbatively, assuming that the off-diagonal
mixing terms are small compared to the diagonal terms.  
Two levels of approximation are considered.  At the first level,
we take $M_Q=M_U=M_D=\msusy$, where $\msusy$ is assumed to be
large compared to $\mz$.  Thus, the radiatively corrected Higgs
mass is determined by $\mha$, $\tanb$, $\msusy$, $A_t$, $A_b$, and
$\mu$.  At the second level, we allow $M_Q$, $M_U$, and $M_D$ to be
arbitrary.   However, as before, we assume that these
soft-supersymmetry-breaking masses are large compared to
$\mz$.  
%Thus, the ratio of off-diagonal to diagonal
%squark squared-mass matrix elements is 
%$\calo(m_t/\msusy)$ and can be treated as an expansion parameter.

At one-loop, the effect of the
squark mixing is to introduce the shifts $\Delta \calmm_{\rm mix}$
and $\left(\Delta\mhpm^2\right)_{\rm mix}$.  For $M_Q=M_U=M_D\equiv\msusy$,
the relevant formulae are given in appendix~B.  For simplicity, we
focus on this case in the remainder of this section.  (For 
the case of arbitrary $M_Q$, $M_U$ and $M_D$, see Appendix C.)
As in Section 2, we note the dependence of $\Delta \calmm_{\rm
mix}(m_t,m_b)$ on the top and bottom quark masses.  If we evaluate $m_t$ 
and $m_b$ at a
suitable scale, then it is possible to account for the dominant effects of
the RG-improvement.  Thus, for non-zero squark mixing, 
eq.~(\ref{simpleform}) is generalized to:
\begin{equation}
\calmm_{\rm 1RG}\simeq\overline{\calmm}_{\rm 1LL}+
\Delta\overline{\calmm}_{\rm mix}\equiv
\calmm_{\rm 1LL}\left(m_t(\mu_t),m_b(\mu_b)\right)+
\Delta\calmm_{\rm mix}\left(m_t(\mu_{\tilde t}),m_b(\mu_{\tilde b})\right)\,,
\label{simplemixform}
\end{equation}
where $\mu_b$ and $\mu_t$ are defined in eq.~(\ref{simpleform}), and
$\mu_{\tilde q}= \msusy$ ($q= t,b)$.  We have extended the overline
notation of Section 2; note in particular that the scales
at which one evaluates $m_t$ and $m_b$ are different in $\calmm_{\rm 1LL}$ and
$\Delta\calmm_{\rm mix}$, respectively.  Intuitively, the squark mixing
correction arises from integrating out heavy squarks that appear
in one-loop corrections to Higgs scalar four-point functions.
As a result, one should choose $\mu_{\tilde q}$ to coincide with
the mass of the heaviest squark.  A more detailed justification
will be given below.
 
Following the discussion of Section 2, we compare the value of
$\mhl$ computed by different procedures.  Prior to RG-improvement, we
first compute $\mhl$ by diagonalizing 
$\calmm_{\rm 1LL}+\Delta\calmm_{\rm mix}$.
Next, we perform RG-improvement as in Ref.~\cite{llog}\
by numerically integrating the RGEs for the Higgs self couplings
starting from supersymmetric boundary conditions, and inserting the
results into the diagonalized tree-level mass matrix.  
Finally, we compare the latter with $\mhl$ computed by diagonalizing
$\overline{\calmm}_{\rm 1LL}+ \Delta\overline{\calmm}_{\rm mix}$ given by
eq.~(\ref{simplemixform}).  These comparisons
are exhibited in a series of figures.  First, we plot
$\mhl$ {\it vs.} $X_t/\msusy$ for $\msusy=\mha=-\mu=1$~TeV for two choices
of $\tanb$ in Fig.~\ref{hhhfig4} [$\tanb=20$] and Fig.~\ref{hhhfig5}
[$\tanb=1.5$]. 
Note that Fig.~\ref{hhhfig4} is of particular interest, since it
allows one to read off the maximal allowed value of $\mhl$ for
$\msusy\leq 1$~TeV.  This maximal value occurs for $|X_t|\simeq 2.4\msusy$.
The reader may worry that this value is too large in light of
our perturbative treatment of the squark mixing.
However, comparisons with exact diagrammatic computations confirm that
these results are trustworthy at least up to the point where the
curves reach their maxima.  From a more
practical point of view, such large values of the mixing are not very
natural; they cause tremendous splitting in the top-squark mass
eigenstates and are close to the region of parameter space where the
SU(2)$\times$U(1) breaking minimum of the
scalar potential becomes unstable relative to color and/or electromagnetic
breaking vacua \cite{casas}.  
%Perhaps it is more prudent to restrict
%$|X_t/\msusy|\lsim 1$ in quoting the maximally allowed $\mhl$.

\begin{figure}[htbp]
\centerline{\psfig{file=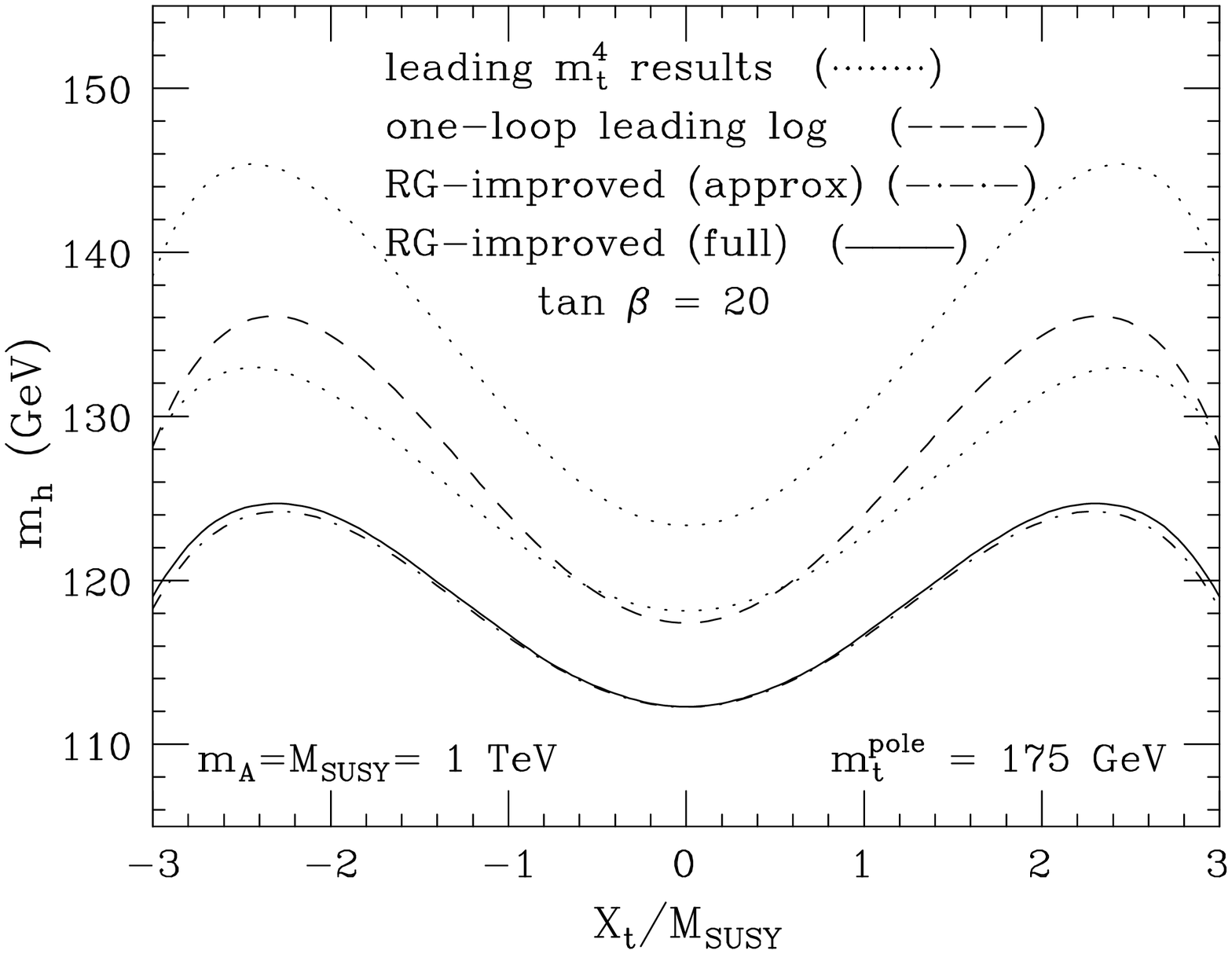,width=12cm,height=9.5cm,angle=90}}
\fcaption{The radiatively corrected light CP-even Higgs mass is plotted
as a function of $X_t/\msusy$, where $X_t\equiv A_t-\mu\cot\beta$, 
for $\msusy=\mha=-\mu=1$~TeV and $\tan\beta=20$. 
See the caption to Fig.~\ref{hhhfig1}.}  
\label{hhhfig4}
%\end{figure}
\vskip1pc
%\begin{figure}[htbp]
\centerline{\psfig{file=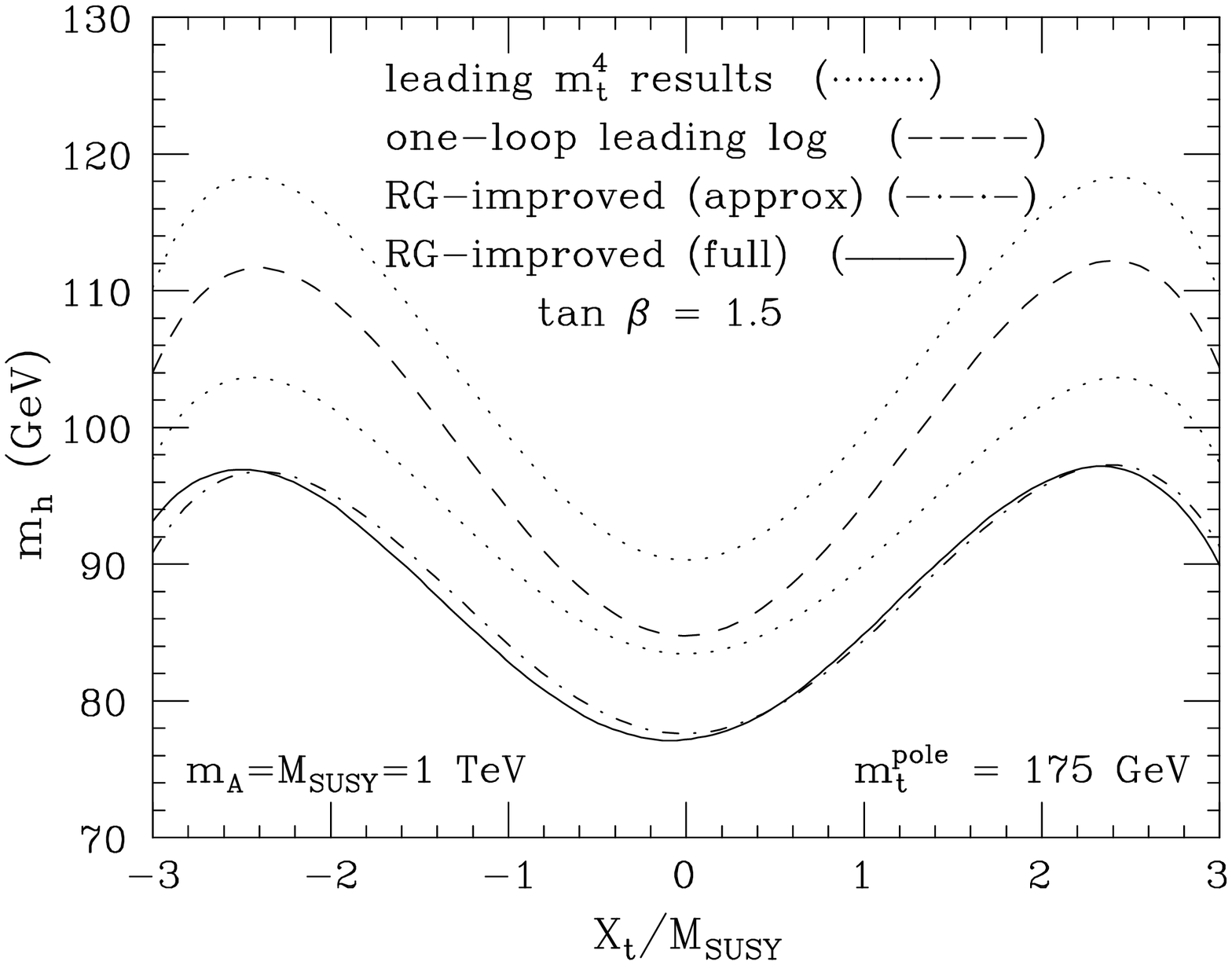,width=12cm,height=9.5cm,angle=90}}
\fcaption{The radiatively corrected light CP-even Higgs mass is plotted
as a function of $X_t/\msusy$, where $X_t\equiv A_t-\mu\cot\beta$, 
for $\msusy=\mha=-\mu=1$~TeV and $\tan\beta=1.5$. 
See the caption to Fig.~\ref{hhhfig1}.}  
\label{hhhfig5}
\end{figure}
 
In Figs.~\ref{hhhfig4} and \ref{hhhfig5}, $\mu=-1$~TeV, {\it i.e.},
as $X_t\equiv A_t-\mu\cot\beta$ varies, so does $A_t$.  
In fact, for $\mha\gg\mz$, the 
dominant one-loop radiative corrections to $\mhl^2$ depend only on $X_t$
and $\msusy$ [see eq.~(\ref{deltamhs})], so that for fixed $X_t$, the
$\mu$ dependence of $\mhl$ is quite weak.  This is illustrated in
Fig.~\ref{hhhfig6}.  For values of $\mha\sim{\cal O}(\mz)$, the $\mu$ 
dependence is slightly more pronounced (although less so for values of
$\tanb\gg 1$) as illustrated in Fig.~\ref{hhhfig7}.     
We also display $\mhl$ as a function of $\msusy$ for a number of
different parameter choices in Fig.~\ref{hhhfig8}.  
In Fig.~\ref{hhhfig9}, we exhibit the $\tanb$ dependence of $\mhl$ for 
two different choices of $X_t$.  Again, we notice that
our approximate formula [eq.~(\ref{simplemixform})], which is depicted by
the dot-dashed line, does remarkably well, and
never differs from the numerically integrated RG-improved value 
(solid line) by more than 1.5~GeV for $\msusy\leq 2$~TeV and $\tanb\geq 1$.

\begin{figure}[htbp]
\centerline{\psfig{file=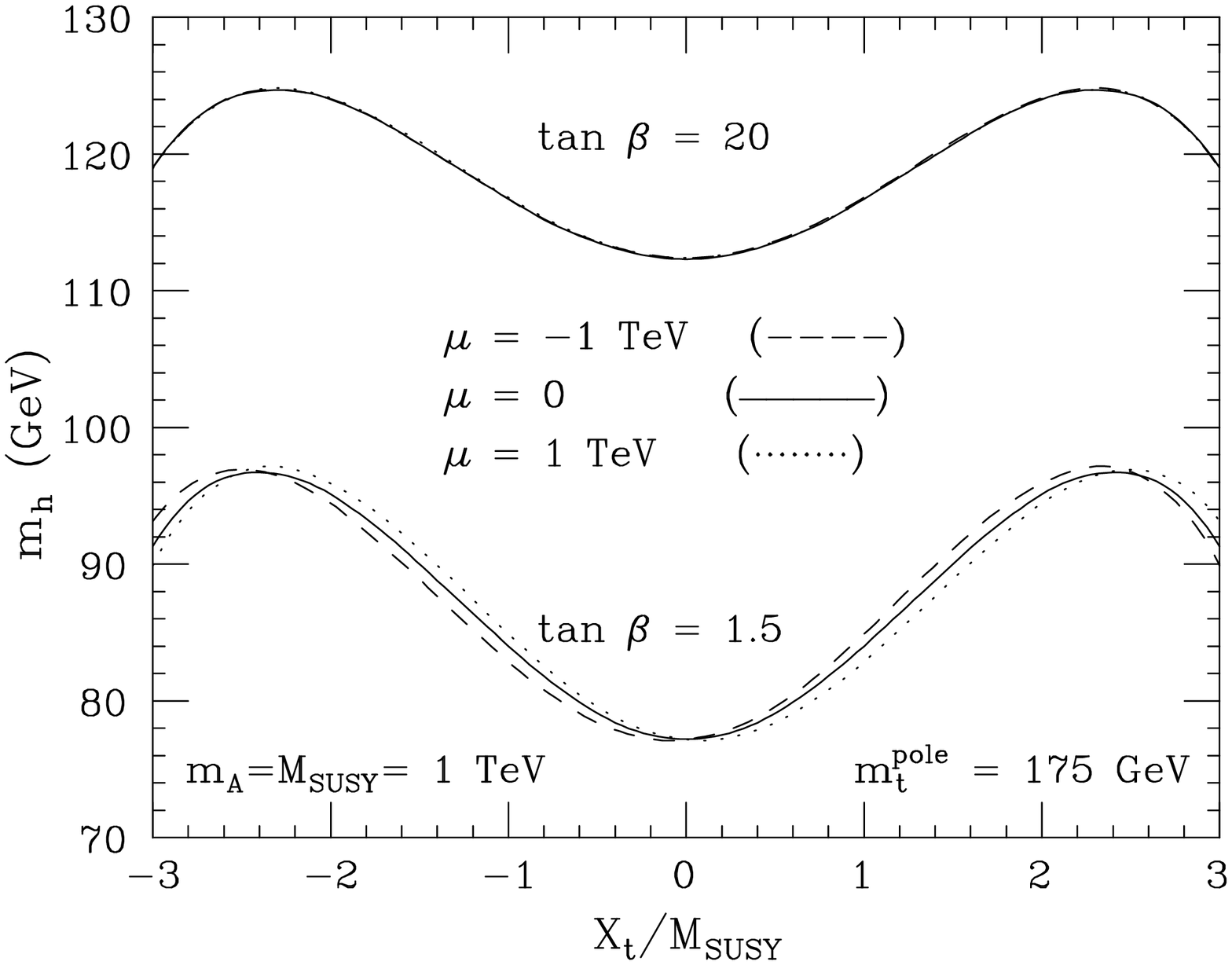,width=12cm,height=9.5cm,angle=90}}
\fcaption{The radiatively corrected, RG-improved 
light CP-even Higgs mass is plotted
as a function of $X_t/\msusy$, where $X_t\equiv A_t-\mu\cot\beta$, 
for $\msusy=\mha=1$~TeV and two choices of $\tanb=1.5$ and 20.
Three values of $\mu$ are plotted in each case: $-1$~TeV [dashed], 0
[solid] and 1~TeV [dotted].  Here, we have assumed that
the diagonal squark squared-masses are degenerate:
$M_Q=M_U=M_D=\msusy$.}
\label{hhhfig6}
%\end{figure}
\vskip1pc
%\begin{figure}[htbp]
\centerline{\psfig{file=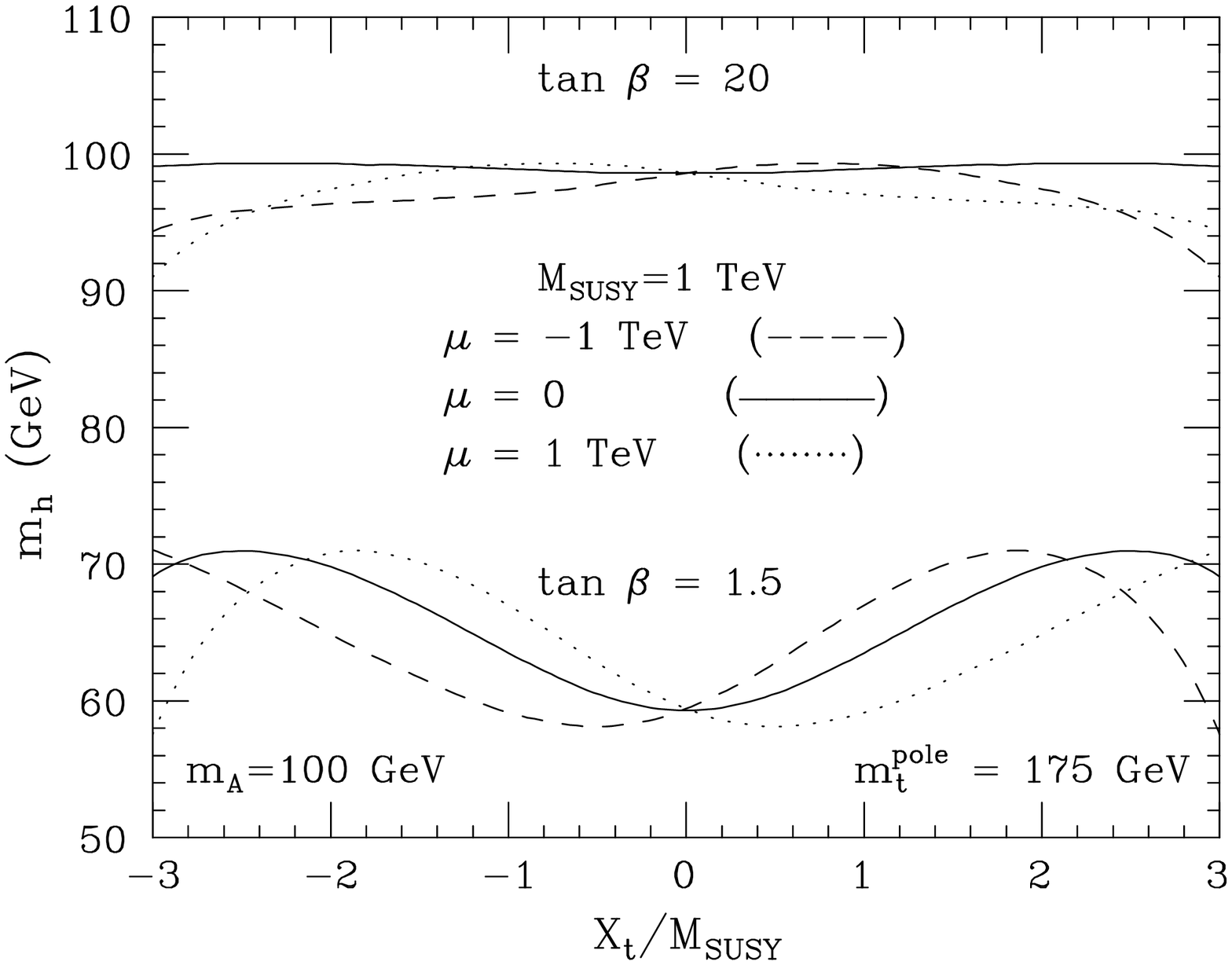,width=12cm,height=9.5cm,angle=90}}
\fcaption{The radiatively corrected, RG-improved 
light CP-even Higgs mass is plotted
as a function of $X_t/\msusy$ for $\msusy=1$~TeV and $\mha=100$~GeV. 
%and two choices of $\tanb=1.5$ and 20.
See the caption to Fig.~\ref{hhhfig6}.}  
\label{hhhfig7}
\end{figure}

\begin{figure}[htbp]
\centerline{\psfig{file=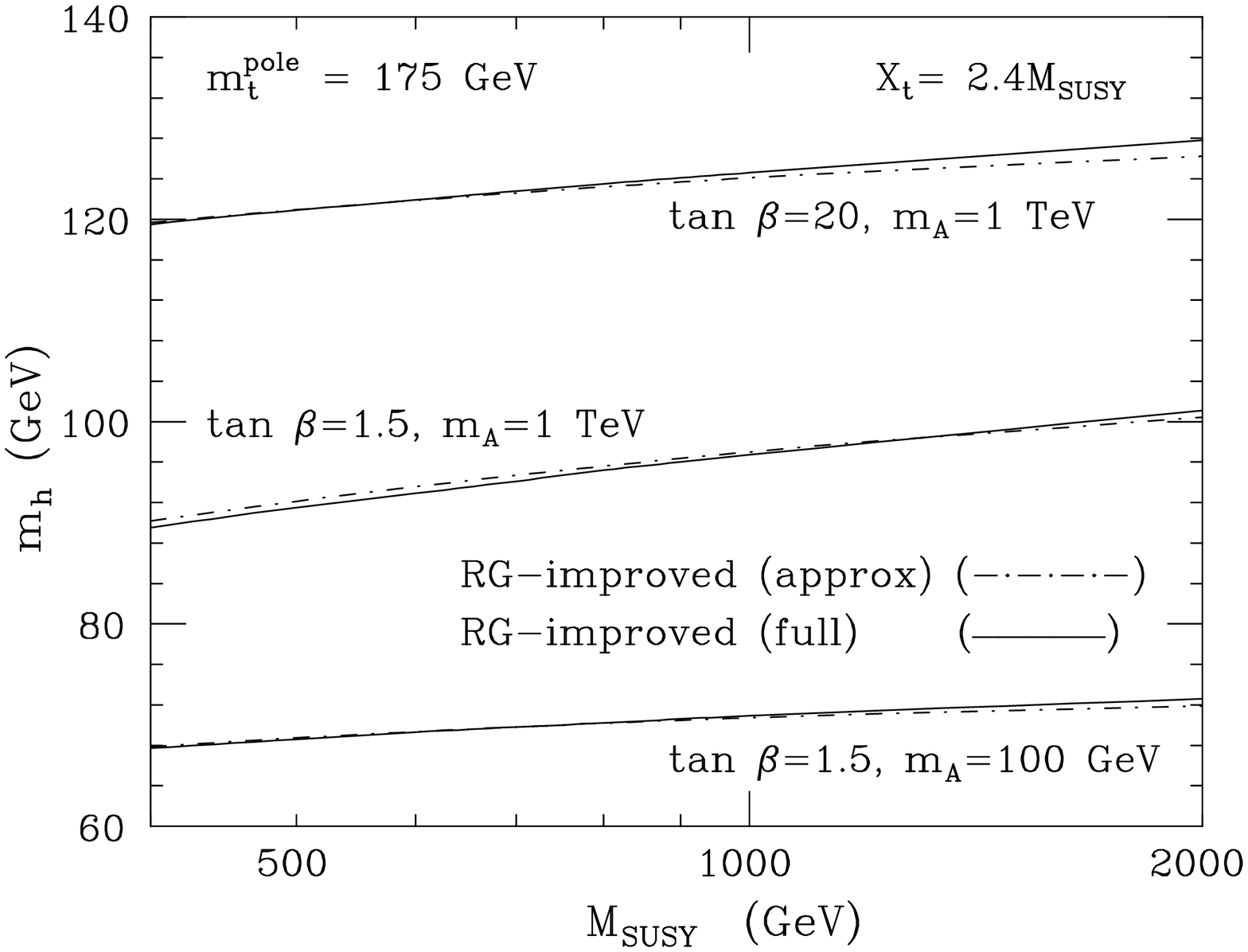,width=12cm,height=9.5cm,angle=90}}
\fcaption{The radiatively corrected, RG-improved 
light CP-even Higgs mass is plotted
as a function of $\msusy$ for $X_t=2.4\msusy$ for three choices of
($\tanb$, $\mha$)= (20,1), (1.5,1), and (1.5,0.1), where $\mha$ is specified
in TeV units.  The solid line depicts the numerically integrated result, 
and the dot-dashed line indicates the result obtained from 
eq.~(\ref{simplemixform}).}
\label{hhhfig8}
%\end{figure}
\vskip1pc
%\begin{figure}[htbp]
\centerline{\psfig{file=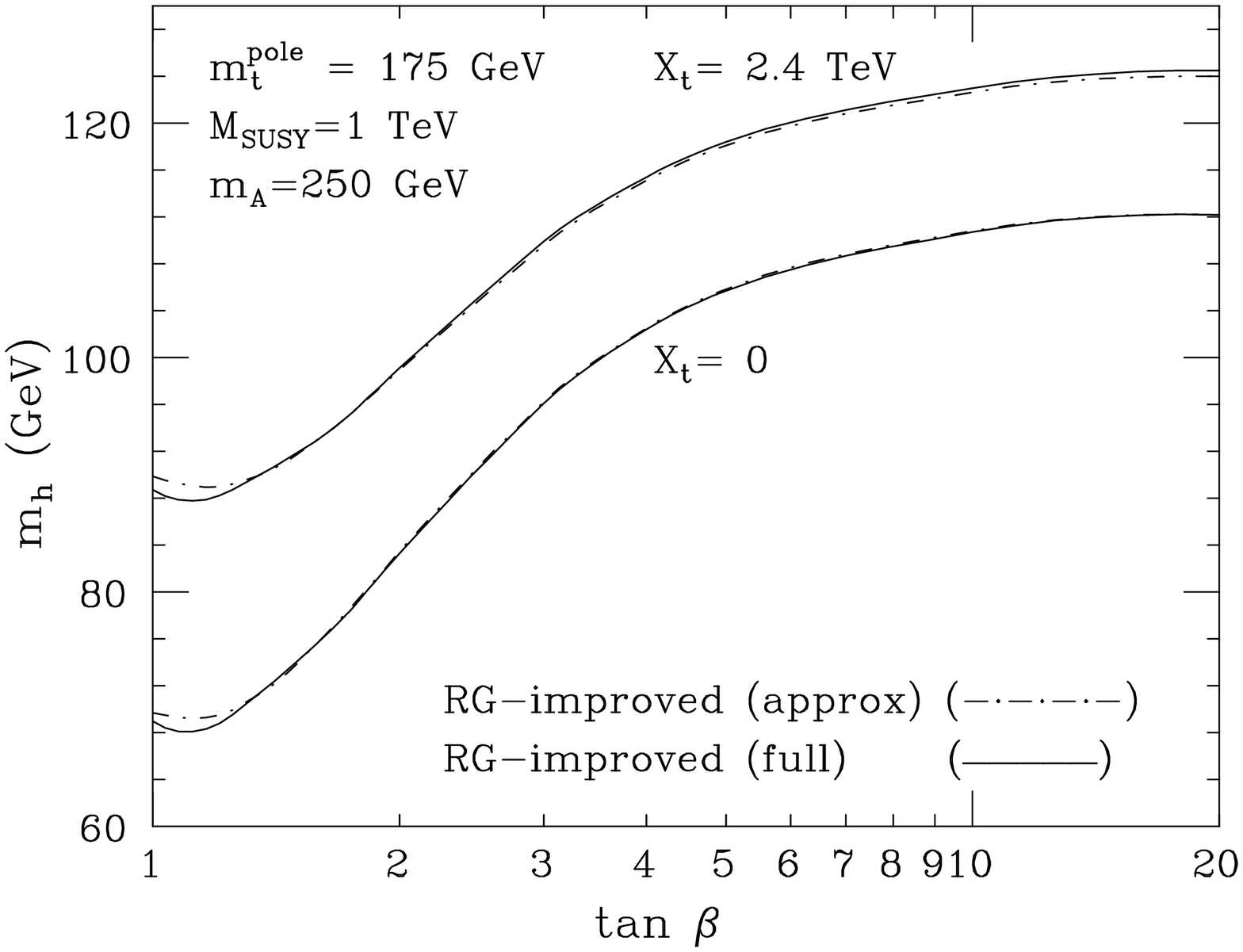,width=12cm,height=9.5cm,angle=90}}
\fcaption{The radiatively corrected, RG-improved 
light CP-even Higgs mass is plotted
as a function of $\tanb$ for $\msusy= 1$~TeV and $\mha= 250$~GeV, for two 
choices of $X_t=0$ and $X_t= 2.4\msusy$.
See the caption to Fig.~\ref{hhhfig8}.}  
\label{hhhfig9}
\end{figure}

Finally, we will present a brief derivation of
eq.~(\ref{simplemixform}). 
As in Section 2, we simplify the analysis by taking $\mha\sim
\calo(\msusy)$,
$m_b=0$, and $\beta=\pi/2$.  To incorporate the effects of top-squark
mixing, eq.~(\ref{lambdaatmsusy}) is modified as follows
\begin{equation}
\lambda(\msusy) = \frac{1}{4}\,\left(g^2+g^{\prime 2}\right) +
  \frac{3A_t^2}{8\pi^2\msusyy}\left[1-{A_t^2\over 12\msusyy}\right]
h_t^4(\msusy)\,.
\label{lambdamixing}
\end{equation}
We may now repeat the steps of Section~2 to obtain:
\begin{equation}
\left(\Delta \mhl^2\right)_{\rm mix}=
 {3g^2m_t^4A_t^2\over8\pi^2\mw^2\msusyy}
\left[1-{A_t^2\over 12\msusyy}\right]
  \left[1+2\left(\gamma_v+
        \frac{\beta_{h_t^2}}{h_t^2}\right)\,
        \ln\left(\frac{\msusyy}{m_t^2}\right)\,\right]\,.
\label{deltahiggsmixing}
\end{equation}
Note the additional factor of 2 multiplying the
logarithmic term as compared with eq.~(\ref{higgsmass1}).  Thus, by
choosing the running top-quark mass at the scale $\mu_{\tilde t}=\msusy$,
one completely absorbs the logarithmic term,
\begin{equation}
\left(\Delta \mhl^2\right)_{\rm mix}=
{3g^2 A_t^2\over8\pi^2\mw^2\msusyy}
\left[1-{A_t^2\over 12\msusyy}\right]m_t^4(\mu_{\tilde t})\,.
\label{deltahiggsmixingabsorbed}
\end{equation}

\section{Discussion and Conclusions}

The masses and couplings of the Higgs bosons of the minimal supersymmetric
model (MSSM) are calculable in terms of two parameters ($\mha$ and
$\tan\beta$) at tree-level.  Including radiative corrections brings in
a dependence on the MSSM particle spectrum, with dominant sensitivity
to the top quark mass and the top squark masses and mixing parameters.
Perhaps the most important prediction of the MSSM Higgs sector
is the value of the lightest
CP-even Higgs mass, $\mhl$, which is predicted to be no greater than
$\mz|\cos 2\beta|+\Delta m$, where $\Delta m$ depends on the size of
the radiative corrections.  This implies that a significant fraction of
the MSSM Higgs parameter space is accessible to LEP-2.  In this regard,
an accurate computation of $\mhl$ is crucial for a reliable assessment
of the capabilities of LEP-2 (as a function of its center-of-mass
energy).  Presently known techniques allow one to compute $\mhl$
with an accuracy of roughly 2 GeV.  If the top squark masses are large
($\gsim 500$~GeV), then it is crucial to include renormalization
group (RG) improvement to obtain the accuracy just quoted.
In this paper, we have presented a simple algorithm for including
RG-improvement in the computation of Higgs masses.  Our proposal
can be applied to either the exact one-loop diagrammatic
computation, or to approximations thereof.  If we denote the one-loop
CP-even Higgs squared-mass matrix by
$\calmm(m_t,m_b,X_t,X_b)$, where we have indicated the dependence
on the third generation quark mass and squark mixing parameters,
then, the RG-improved squared-mass matrix is well approximated by
\begin{equation}
\overline{\calmm}\equiv
  \calmm\left[m_t(\mu_{\tilde t}),m_b(\mu_{\tilde b}),X_t,X_b\right]
-\calmm\left[m_t(\mu_{\tilde t}),m_b(\mu_{\tilde b}),0,0\right]
+\calmm\left[m_t(\mu_t),m_b(\mu_b),0,0\right]
\label{rgimprovedmass}
\end{equation}
This summarizes the results of eq.~(\ref{simpleform}) and 
eq.~(\ref{simplemixform}).  Note that
eq.~(\ref{rgimprovedmass}) can be applied even in the case where the 
effects of squark mixing are not explicitly separated out.
As an example, consider the full diagrammatic one-loop computation, whose
result we shall denote by $\calmm_{\rm 1LC}$.  To carry out the
RG-improvement, one would 
subtract off the one-loop leading logarithmic pieces, $\calmm_{\rm 1LL}$,
and replace it with its RG-improved value, $\calmm_{\rm 1RG}$.  If we 
implement the RG-improvement using our analytic method described by
eq.~(\ref{rgimprovedmass}), we obtain     
\begin{equation}
\overline{\calmm}_{\rm 1LC}\equiv\calmm_{\rm 1LC}+\overline{\calmm}_{\rm 1LL}
+\Delta\overline{\calmm}_{\rm mix}-\calmm_{\rm 1LL}-\Delta\calmm_{\rm mix}\,.
\label{accurate}
\end{equation}
In cases that we have checked, we find that 
$\calmm_{\rm 1LC}-\calmm_{\rm 1LL}-\Delta\calmm_{\rm mix}$
produces no more than about a 1~GeV shift in the predicted Higgs masses.  
Thus, we expect that the numerical results obtained in this paper will shift
by no more than $\sim 1$~GeV if non-leading logarithmic terms are included.

We find that if our algorithm is applied to the leading log one-loop
corrections (given in Appendices A and C), plus the leading terms
resulting from squark mixing (given in Appendices B and C), 
%the approximate RG-improved formula yields numerical results for
%$\mhl$ that are accurate to within $\sim 2$~GeV for supersymmetric
%particle masses below 2~TeV.  
we reproduce the full (numerically integrated) RG-improved value of
$\mhl$, to within an accuracy of less than 2~GeV (assuming that 
supersymmetric particle masses lie below 2 TeV).  Although we have not
focused our attention to the heavier Higgs states, we assert that our 
algorithm also yields accurate results for the mass of the
heavier CP-even Higgs boson, $\mhh$.  Our approximation to the radiatively
corrected charged Higgs mass is slightly less accurate only because the
leading $m_t$ enhanced terms are not as dominant as in the neutral Higgs
sector.\footnote{The approximation to the radiatively 
corrected charged Higgs mass
can be improved by including sub-dominant terms not contained in the 
formulae given in this paper; see Ref.~\cite{madiaz} for further details.} 
Thus, the methods described in this paper provide a
simple and powerful technique for obtaining an accurate determination
of the radiatively corrected MSSM Higgs masses and CP-even mixing angle.

\section*{Acknowledgments}

We gratefully acknowledge many fruitful interactions with Marcela Carena,
Mariano Quiros, and Carlos Wagner.  We also appreciate useful conversations
with Damien Pierce and Fabio Zwirner.  Two of us (HEH and RH) would like to
thank the Institute for Theoretical Physics in Santa Barbara for their partial
support under the National Science Foundation Grant No. PHY94-07194
during the final phase of this project.  One of us (HEH) is grateful
for the hospitality of the Institut f\"ur Theoretische Teilchenphysik in 
Karlsruhe and the  Max-Planck-Institut f\"ur Physik in Munich, 
where some of this work was accomplished.  HEH is
supported in part by the U.S. Department of Energy, and AHH is supported
by the Graduiertenkolleg {\it Elementarteilchenphysik}, Karlsruhe. 

%**********************************************************************
\clearpage
%\vspace{1.5cm}
\appendixA{Appendix A:
One-Loop Leading Logarithmic Corrections to the MSSM Higgs Masses}

 \noindent
We suppose that the supersymmetric particle spectrum is characterized by
one mass scale, called $\msusy$ (assumed to be larger than $m_t$)  
Specifically, we assume that
terms proportional to $\ln(\msusyy/M_X^2)$, where $M_X$ stands for
the mass of any supersymmetric particle, are much smaller than terms
proportional to 
$\ln(\msusy^2/\mz^2)$ and can be neglected.\footnote{For a widely spread
supersymmetric particle 
spectrum where this approximation is inappropriate,
these corrections are listed in Appendices~B and C (squarks and sleptons)
and D (charginos and neutralinos).}
This assumption is made to keep 
the formulae that follow relatively simple.  In fact, one can account
for some spread in the supersymmetric mass spectrum by interpreting
$\msusy$ appropriately depending on where it occurs in the formulae
below.  In this case additional correction terms arise, which are
treated in Appendices~C and~D. 

In the one-loop leading logarithmic approximation,
the CP-even Higgs squared-mass matrix is given by:
\begin{equation} \label{llcpevenhiggs}
\calmm_{\rm 1LL}=\calmm_0+\Delta\calmm_{1LL}\,,
\end{equation}

\noindent
where $\calmm_0$ is the tree level result given in eq.~(\ref{mssmtree}), and

\begin{eqnarray} \label{mthree}
(\Delta\calm_{11}^2)_{\rm 1LL} & = &
{g^2\mzz\cbb\over96\pi^2\cw2}\Big[
P_t~\ln\left({\msusyy\over m_t^2}\right)\nonumber \\[3pt]
&&~~+\Big(12N_c{m_b^4\over\mz^4\cbiv}-6N_c{m_b^2\over\mzz\cbb}
+P_b+P_f+P_g+P_{2H} \bigg)\ln\left({\msusyy\over
m_Z^2}\right)\nonumber\\[3pt]
&&~~+\Theta(\mha-\mz)
(P_{1H}-P_{2H})\ln\left({\mha^2\over\mz^2}\right)\Big]\,,\nonumber \\[6pt]
(\Delta\calm_{22}^2)_{\rm 1LL} & = &
{g^2\mzz\sbb\over96\pi^2\cw2}\Big[\bigg(P_b+P_f+P_g+P_{2H}
\bigg)\ln\left({\msusyy\over m_Z^2}\right)\nonumber \\[3pt]
&&~~+\bigg(12N_c{m_t^4\over\mz^4\sbiv}-6N_c{m_t^2\over\mzz\sbb}
+P_t\bigg)\ln\left({\msusyy\over m_t^2}\right)\nonumber \\[3pt]
&&~~+\Theta(\mha-\mz)
(P_{1H}-P_{2H})\ln\left({\mha^2\over\mz^2}\right)
+2N_c{\mt^2\over\mz^2\sbiv}\Big]\,, \nonumber \\[6pt]
(\Delta\calm_{12}^2)_{\rm 1LL} & = &
-{g\mzz\sb\cb\over96\pi^2\cw2}\bigg[\bigg(P_t-3N_c{m_t^2\over\mzz\sbb}
\bigg)\ln\left({\msusyy\over m_t^2}\right)\nonumber\\[3pt]
&&~~+\Big(P_b-3N_c{m_b^2\over\mzz\cbb}+P_f+P_g'+P_{2H}'
\bigg)\ln\left({\msusyy\over
m_Z^2}\right)\nonumber\\[3pt]
&&~~-\Theta(\mha-\mz)
(P_{1H}+P^\prime_{2H})\ln\left({\mha^2\over\mz^2}\right)\Big]\,.
\end{eqnarray}
\clearpage 
\noindent
In eq.~(\ref{mthree}), the constants $P_i$ are defined as
\begin{eqnarray}
P_t & \equiv & N_c(1-4 e_t \sw2+8 e_t^2 \swiv)\,,\nonumber\\
P_b & \equiv & N_c(1+4 e_b \sw2+8 e_b^2 \swiv)\,,\nonumber\\
P_f & \equiv & N_c(N_g-1)\left[\,2-4\sw2+8(e_t^2+e_b^2)\swiv\,\right]
+N_g\left[\,2-4 \sw2+8 \swiv\right]\,,\nonumber\\
P_g & \equiv & - 44 + 106 \sw2 - 62 \swiv\,,\nonumber \\
P_g'& \equiv & 10 + 34 \sw2 - 26 \swiv\,,\nonumber \\
P_{2H} & \equiv & - 10 + 2 \sw2 - 2 \swiv\,,\nonumber \\
P_{2H}'& \equiv & 8 - 22 \sw2 + 10 \swiv\,,\nonumber \\
P_{1H} & \equiv &
  -9 \cos^4 2\beta + \left( 1 - 2 \sw2 + 2 \swiv\right)\cos^2 
2\beta\,,
\label{defpp}
\end{eqnarray}
with the number of colors, $N_c=3$, the number of generations, $N_g=3$,
and the quark charges given by $e_t\equiv 2/3$ and $e_b\equiv -1/3$.
The origin of the various terms above can be found in Ref.~\cite{llog}.
The most important contribution to the one-loop radiative corrections above
arises from the term in $\Delta\calmm_{22}$ that is proportional to $\mt^4$.
If one discards all other one-loop radiative corrections, then
one obtains $\Delta\calmm_{\rm 1LT}$ given in eq.~(\ref{topapprox}).

In principle, the two cases $\mha\lsim\calo(\mz)$ and $\mha\gg\mz$
must be treated separately.  However, by including 
the terms proportional to $\ln(\mha^2/\mz^2)$ 
in eq.~(\ref{mthree}),\footnote{Strictly
speaking, these terms are meaningful for $\mz\ll\mha\leq\msusy$.  The
step function in eq.~(\ref{mthree}) implies
that these terms should not be included if
$\mha\leq\mz$.}  we interpolate between the small and large $\mha$
cases.  This yields the correct leading log expression for $\mhl$.
The large $\mha$ analysis is more conveniently done in a basis where
$\hl$ and $\hh$ are approximate eigenstates.  If ${\cal R}$ is the
rotation matrix which rotates a 2-vector counterclockwise by angle 
$\beta$, then ${\cal R}^{-1}\calmm{\cal R}$ yields 
\begin{equation} 
\pmatrix{\mhl^2&m^2_{hH}\cr m^2_{hH}&\mhh^2}\simeq
\pmatrix{\calmm_{11}\cbb+\calmm_{22}\sbb+\calmm_{12}\sin2\beta &
\half(\calmm_{22}-\calmm_{11})\sin2\beta+\calmm_{12}\cos2\beta\cr
\half(\calmm_{22}-\calmm_{11})\sin2\beta+\calmm_{12}\cos2\beta &
\calmm_{11}\sbb+\calmm_{22}\cbb-\calmm_{12}\sin2\beta\cr}\,,
\label{largemacase}
\end{equation}
where the identification of the diagonal elements above follows from 
the fact that $m^2_{hH}=\calo(\mz^2)$ in the limit of $\mha\gg\mz$.
Using the above results, it follows that for $\mha\gg\mz$,
\begin{eqnarray}
(\mhl^2)_{\rm 1LL} &=& \mzz\cos^2 2\beta
+{g^2m_Z^2\over96\pi^2\cw2}
\Bigg\{\bigg[12N_c{m_b^4\over\mz^4}-6N_c{m_b^2\over\mzz}\ctwob
+P_f\ctwobb \nonumber \\
&&~~+(P_{g}+P_{2H})(\sbiv+\cbiv)
 -2(P_{g}'+P_{2H}')\sbb\cbb
 \bigg]\ln\left({\msusyy\over\mzz} \right)\nonumber \\
&&~~+\bigg[12N_c{m_t^4\over\mz^4}+
 6N_c{m_t^2\over\mzz}\ctwob+P_t\ctwobb\bigg]
 \ln\left({\msusyy\over m_t ^2}\right)\nonumber \\
&&~~-\bigg[P_{2H}\left(\cbiv+\sbiv\right)-2 P_{2H}'\cbb\sbb -P_{1H}\bigg]
 \ln\left({\mha^2\over\mzz}\right)\Bigg\}\,,\nonumber \\
(\mhh^2)_{\rm 1LL}&=&\mha^2+\mz^2\sin^2 2\beta+\calo(g^2\mz^2)\,,\nonumber \\
(m^2_{hH})_{\rm 1LL}&=&-\mz^2\sin2\beta\cos2\beta+\calo(g^2\mz^2)\,.
\label{mhltot}
\end{eqnarray}
In order to compare the results of our analytic
approximation to the RG-improved Higgs squared-mass [eq.~(\ref{simpleform})],
one must integrate the Higgs self-couplings of the two-Higgs-doublet model
from $\msusy$ down to $\mha$, match onto the one-Higgs-doublet
model, and finally evolve the Standard Model Higgs self-coupling from
$\mha$ down to $\mz$.  However, we note that the contribution of the terms 
proportional to $\ln(\mha^2/\mzz)$
are quite small, leading to a mass shift in $\mhl$ of no more than 1 GeV
(for $\mha\leq 1$~TeV). 
As a result, we simplified the analysis in generating the graphs shown
in Section 2 and 3 by omitting the $\mha$ dependence in the  
radiative corrections.\footnote{In producing the graphs shown in Sections 2
and 3, we also have omitted the non-leading logarithmic term proportional to
$m_t^2$ that appears in $\Delta\calmm_{22}$, since this term is not
picked up by the numerical integration of the RGEs.}
That is, we omitted all terms proportional
to $\ln(\mha^2/\mzz)$ that appear in the formulae above.  We then
compared the analytic approximation so obtained 
with the two-Higgs-doublet model numerically 
integrated from $\msusy$ down to $\mz$, with no matching onto the
one-Higgs-doublet model as described above.  In this approximation, the
dependence of the Higgs masses on $\mha$ enters only through the
tree-level expressions.  The impact of the $\mha$ dependence of the
radiative corrections will be discussed briefly at the end of Appendix C. 

Although the one-loop leading log terms in $\mhl^2$ are correctly reproduced
by using the results given in eqs.~(\ref{mthree})--(\ref{mhltot}),
the analogous terms in $\mhh^2$ obtained by the above procedure
do not catch all terms proportional to $\ln(\mha^2/\mz^2)$.
The reason is related to the fact that below the scale $\mha$, the
effective low-energy Higgs sector consists of just one scalar,
$\hl$.  Thus, for example, for $\mha=\msusy$
one would find $\mhh^2=\mha^2+\mz^2\sin^2 2\beta$, where the 
parameters on the right hand side should be considered as running 
parameters evaluated at the scale $\mha$ (and similarly for
$m^2_{hH}=-\mz^2\sin2\beta\cos2\beta$).  Then, the correct
leading log pieces in the unspecified $\calo(g^2\mz^2)$ terms 
in eq.~(\ref{mhltot}) would be
obtained once the running parameters were expressed in terms of
physical parameters.  We shall not provide explicit expressions
for the resulting $\calo(g^2\mz^2)$ terms, since these 
corrections\footnote{Note that the above arguments imply that 
if $\mha=\msusy$ then the leading $m_t^4$ contribution to the
$\calo(g^2\mz^2)$ terms is absent, while for $\mt\leq\mha\leq\msusy$, the 
leading $m_t^4$ term is proportional to $m_t^4\ln(\msusyy/\mha^2)$.}
yield only a small fractional shift to $\mhh^2$.  One subtle point
involves the definition of $\tanb$.  The angle $\beta$ that appears
in the expression for $\mhl^2$
in eq.~(\ref{mhltot}) is implicitly defined at the scale $\mz$.
However, for $\mha\gg\mz$, the Higgs sector is effectively the
one-doublet scalar sector of the Standard Model below the scale
$\mha$.  Thus, another choice is to define $\tanb$ as the ratio
of vacuum expectation values evaluated at $\mha$.  To convert
between definitions is a simple task.  In the expression for $\mhl^2$
given in eq.~(\ref{mhltot}), it suffices to replace the 
tree-level contribution, $\mz^2\cos^2 2\beta(\mz)$, using the leading log
expression \cite{llog}
\begin{equation}
\cos^2 2\beta(\mz)=\cos^2 2\beta(\mha)+{g^2 N_c\cos 2\beta\over
8\pi^2\mw^2}\left[m_t^2\cbb\ln\left({\mha^2\over\mt^2}\right)
-m_b^2\sbb\ln\left({\mha^2\over\mz^2}\right)\right]\,.
\label{tanbrun}
\end{equation}
However, we alert the reader that we have not made this replacement
in any of our formulae.  If not explicitly indicated,
all quoted formulae in this paper 
assume that $\tan\beta$ is defined at $\mz$.
%unless otherwise explicitly indicated,

We can improve the above formulae by reinterpreting the meaning of
$\msusy$.  For example, all terms proportional to $\ln(\msusy^2/\mt^2)$
arise from diagrams with loops involving the top quark and top-squarks.
Explicit diagrammatic computations then show that
we can reinterpret $\msusy^2=M_{\tilde t_L}M_{\tilde t_R}$.
Note that with this reinterpretation of $\msusyy$, the top quark and top
squark loop contributions to the Higgs masses cancel exactly
when $M_{\tilde t_L}=M_{\tilde t_R}=m_t$, as required
in the supersymmetric limit.  
Likewise, in terms proportional to $P_b$ or powers of $m_b$ multiplied by
$\ln(\msusyy/\mzz)$, we may reinterpret $\msusy=M_{\tilde b_L}M_{\tilde b_R}$.
Terms proportional to $P_f\ln(\msusyy/\mzz)$ 
come from loops of lighter quarks and
leptons (and their supersymmetric partners) in an obvious way, and
the corresponding $\msusyy$ can be reinterpreted accordingly.
Additional contributions arising in the case of non-degenerate squark
masses are treated in Appendix C.
The remaining leading logarithmic terms arise from gauge and
Higgs boson loops and their supersymmetric partners.  The best we can
do in the above formulae is to interpret $\msusy$ as an average
neutralino and chargino mass.  To incorporate thresholds more precisely
requires a more complicated version of eq.~(\ref{mthree}), 
which can be easily derived from formulae given in ref.~\cite{llog}.  
The corresponding expressions are summarized in Appendix D.  However,
%But, to incorporate such fine
%details is not very useful, since the change in the computed Higgs masses
%is numerically inconsequential.  Moreover, the terms omitted
%due to the neglect of the details of the neutralino and chargino
%spectrum 
the impact of these corrections are no more important than the
non-leading logarithmic terms which have been discarded.  
The most significant
non-logarithmic correction is one term that scales with $m_t^2$ and thus has
been included in eq.~(\ref{mthree}).  
This is surely the largest of such corrections.
One can check that it yields at most a 1~GeV shift in the computed Higgs
masses.

Finally, we give the one-loop leading logarithmic expression for the
charged Higgs mass.  First, assuming that $\mha\lsim\calo(\mz)$, 
\begin{eqnarray}
(m_{H^{\pm}}^2)_{\rm 1LL}\!\!&=&\!\!
\mha^2+\mw^2 + {{N_c g^2}\over{32\pi^2m_W^2}}
\Bigg[{{2m_t^2m_b^2}\over{\sbb\cbb}}-m_W^2
\bigg({{m_t^2}\over{\sbb}}+{{m_b^2}\over{\cbb}}\bigg)
+{\textstyle{2\over 3}}m_W^4\Bigg]
\ln\left({{\msusy^2}\over{m_t^2}}\right)\nonumber \\
&&\qquad\qquad\qquad+{g^2{m_W^2}\over{48\pi^2}}\left[N_c(N_g-1)+N_g-9
+15\tan^2\theta_W\right]
\ln\left({{\msusy^2}\over{m_W^2}}\right)\,.
\label{llform}
\end{eqnarray}
For $\mha\gg\mz$,
additional terms arise proportional to $\ln(\mha^2/\mw^2)$.  
For example, for $\mha=\msusy$, the leading logarithmic corrections
can be obtained from $\mhpm^2=\mha^2+\mw^2$, 
where $\mw^2$ is treated as a running parameter evaluated at $\mha$.
Re-expressing $\mw(\mha)$ in terms of the physical $W$ mass yields
the correct one-loop leading log corrections
for $\mz\leq\mha\leq\msusy$.  Again, we omit the
explicit expressions since these corrections generate only a small
relative shift to the heavy charged Higgs mass.

A good approximation to the RG-improved Higgs squared-mass 
corrections is implemented according to the algorithm of Section 2
by replacing $m_t$ and $m_b$ in the above formulae by the
corresponding running parameters evaluated at $\mu_t=\sqrt{\mt\msusy}$
and $\mu_b=\sqrt{\mz\msusy}$, respectively.

%\par
%\vspace{1cm}\noindent
%{\bf
\appendixB{Appendix B:
Leading Squark Mixing Corrections to the MSSM Higgs Masses}
\noindent
When squark mixing effects are taken into account, the CP-even 
Higgs squared-mass matrix and $m_{\hpm}^2$ receive additional one-loop
corrections beyond those given in Appendix A.  We need only consider
the effect of mixing among the third-generation squarks (since all 
such mixing effects are proportional to the corresponding quark 
mass).  In order to keep the formulae simple, we assume that the
diagonal elements of the $2\times 2$ squark squared-mass matrices 
[eqs.~(\ref{stopmatrix}) and (\ref{sbotmatrix})]
are degenerate and given by $\msusyy$.  (In Appendix C, we treat the
more complex case of non-universal squark squared-masses.)
It is convenient to define
\begin{eqnarray}
X_t&\equiv&A_t-\mu\cot\beta\,,\qquad\qquad
Y_t\equiv A_t+\mu\tan\beta\,,\nonumber \\
X_b&\equiv&A_b-\mu\tan\beta\,,\qquad\qquad
Y_b\equiv A_b+\mu\cot\beta\,.
\label{xdefs}
\end{eqnarray}
We assume that the mixing terms $\mt X_t$ and $\mb X_b$ are not too 
large.\footnote{Formally, the expressions given in this Appendix
are the results of an expansion in the variable $(M_1^2-M_2^2)/
(M_1^2+M_2^2)$, where $M_1^2$, $M_2^2$ are the squared-mass eigenvalues
of the squark mass matrix.   Thus, we demand that $\mt X_t/\msusyy\ll 
1$.  For example, for $\msusy=1$~TeV, values of $X_t/\msusy\lsim 3$
should yield an acceptable approximation based on the formulae
presented here.}
Then, the elements of the CP-even Higgs squared-mass matrix
are given by:
\begin{equation} \label{fullcorr}
\calmm=\calmm_{\rm 1LL}+\Delta\calmm_{\rm mix}\,,
\end{equation}
where $\calmm_{\rm 1LL}$ has been given in 
eqs.~(\ref{llcpevenhiggs}) and (\ref{mthree}), and
\begin{eqnarray} \label{deltacalms}
(\Delta\calmm_{11})_{\rm mix}&=& {g^2N_c\over 32\pi^2\mw^2\msusy^2}\Biggl[
{4m_b^4A_b X_b\over\cbb}\left(1-{A_b X_b\over 12\msusy^2}\right)
-{m_t^4\mu^2 X_t^2\over 3\msusy^2\sbb}\nonumber \\
&&\qquad\qquad -\mz^2 m_b^2A_b(X_b+\third A_b)-\mz^2
m_t^2\mu\cot\beta(X_t+\third\mu\cot\beta)\Biggr]\,,\nonumber \\
(\Delta\calmm_{22})_{\rm mix}&=& {g^2N_c\over 32\pi^2\mw^2\msusy^2}\Biggl[
{4m_t^4A_t X_t\over\sbb}\left(1-{A_t X_t\over 12\msusy^2}\right)
-{m_b^4\mu^2 X_b^2\over 3\msusy^2\cbb}\nonumber \\
&&\qquad\qquad -\mz^2 m_t^2A_t(X_t+\third A_t)-\mz^2
m_b^2\mu\tan\beta(X_b+\third\mu\tan\beta)\Biggr]\,,\nonumber \\
(\Delta\calmm_{12})_{\rm mix}&=& {-g^2N_c\over 64\pi^2\mw^2\msusy^2}\Biggl[
{4m_t^4\mu X_t\over\sbb}\left(1-{A_t X_t\over 6\msusy^2}\right)
+{4m_b^4\mu X_b\over\cbb}\left(1-{A_b X_b\over 6\msusy^2}\right)\nonumber \\
&&\  -\mz^2 m_t^2\cot\beta[X_t Y_t+\third(\mu^2+A_t^2)]
-\mz^2 m_b^2\tan\beta[X_b Y_b+\third(\mu^2+A_b^2)]\Biggr]\,.
\end{eqnarray}
%If the diagonal $b$-squark and $t$-squark masses differ, then it is a 
%simple matter to correct the above formulae by identifying $\msusyy$
%appropriately, depending on whether the corresponding term is
%proportional to $m_b$ or $m_t$.  If the diagonal elements of the
%$t$-squark (or $b$-squark) mass matrix are not degenerate, the above
%formulae may still be useful. Explicit diagrammatic computations show
%that $\msusyy$ is equal to the {\it average} of the two diagonal
%squark squared-masses.  However, in this case, eq.~(\ref{deltacalms})
%misses terms proportional to $\Delta M_{\widetilde Q}^2/\msusyy$, where
%$\Delta M_{\widetilde Q}^2$ is the difference of diagonal squared-masses.
%If $\Delta M_{\widetilde Q}^2\simeq\calo(\mz^2)$ [which might be 
%expected; see eqs.~(\ref{stopmatrix} and (\ref{sbotmatrix})], 
%then one should also neglect the terms in
%eq.~(\ref{deltacalms}) proportional to $\mz^2$.  Still, the terms
%proportional to $m_t^4$ would presumably dominate (terms proportional
%to $m_b^4$ would be important only if $\tanb\gg 1$). 
%We have checked that the corrections to Higgs masses from 
%eq.~(\ref{deltacalms}) provides a fairly accurate approximation over 
%a large range of supersymmetric parameters as long as $\Delta 
%M_{\tilde Q}$ or the size of the off-diagonal squark squared-mass 
%terms are not too large.

If $\mz\ll\mha\leq\msusy$,
%These corrections do not depend on the value of $\mha$ (although there
%is an implicit assumption that $\mha\leq\msusy$).  If $\mha=\msusy\gg\mz$,
then it is again convenient to rotate the neutral scalar Higgs basis
as discussed in Appendix A.  One then finds that the Higgs squared-masses
obtained in eq.~(\ref{mhltot}) are shifted by the following expressions
\begin{eqnarray} \label{deltamhs}
(\Delta\mhl^2)_{\rm mix}\!\!&=&\!\!{g^2 N_c\over 16\pi^2\mw^2\msusyy}
\Biggl\{2m_t^4 X_t^2\left(1-{X_t^2\over
12\msusyy}\right)+2m_b^4 X_b^2\left(1-{X_b^2\over
12\msusyy}\right) \nonumber \\
&&\quad+\half\mz^2\cos2\beta\left[m_t^2
\left(X_t^2+\third(A_t^2-\mu^2\cot^2\beta)\right)
-m_b^2\left(X_b^2+\third(A_b^2-\mu^2\tan^2\beta)\right)\right]
\Biggr\}\,, \nonumber\\
(\Delta\mhh^2)_{\rm mix}\!\!&=&\!\!{g^2 N_c\over 16\pi^2\mw^2\msusyy}
\Biggl\{2m_t^4X_t Y_t\cot^2\beta\left(1-{X_t Y_t\over 12\msusyy}\right)
+2m_b^4X_b Y_b\tan^2\beta\left(1-{X_b Y_b\over 12\msusyy}\right)
\nonumber \\
&&\quad -\mz^2\left[m_t^2\cbb\left(X_t Y_t+\third(A_t^2+\mu^2)\right)
+m_b^2\sbb\left(X_b Y_b+\third(A_b^2+\mu^2)\right)\right]\Biggr\}\,.
\end{eqnarray}

Squark mixing effects
also lead to modifications of the charged Higgs squared-mass. 
One finds that the charged Higgs squared-mass
obtained in eq.~(\ref{llform}) is shifted by
\begin{eqnarray} \label{deltamhpm}
(\mhpm^2)_{\rm mix} \!\!&=&\!\! 
 {N_cg^2\over192\pi^2\mw^2\msusy^2}\Biggl[
 {2\mt^2\mw^2(\mu^2-2A_t^2)\over\sbb}
  +{2\mb^2\mw^2(\mu^2-2A_b^2)\over\cbb} \nonumber \\
&& -3\mu^2\left({m_t^2\over\sbb}
  +{m_b^2\over\cbb}\right)^2 +{\mt^2\mb^2\over\sbb\cbb}\left(3(A_t+A_b)^2
-{(A_t A_b -\mu^2)^2\over\msusyy}\right)\Biggr]\,.
\end{eqnarray}

A good approximation to the RG-improved Higgs squared-mass 
corrections is implemented according to the algorithm of Section 3
by replacing $m_t$ and $m_b$ in the above formulae by the
corresponding running parameters evaluated at $\mu=\msusy$.

%\par
%\vspace{1cm}\noindent
%{\bf
\appendixC{Appendix C:
Non-Universal Squark Mass Corrections}
\noindent
The formulae of Appendices A and B were obtained under the
assumption that $M_Q=M_U=M_D\equiv\msusy$, where $M_Q$, $M_U$,
and $M_D$ are the diagonal soft-supersymmetry-breaking squark
mass parameters defined in eqs.~(\ref{stopmatrix}) and (\ref{sbotmatrix}).
In this Appendix,
we generalize the results of the previous two appendices to
allow for unequal diagonal scalar masses.
 
First, the results of Appendix A are modified as follows.  For
the neutral CP-even squared-mass matrix, replace in
eqs.~(\ref{mthree})--(\ref{mhltot}) all occurrences of
$\ln(\msusyy/\mt^2)$ with $\ln(M_Q M_U/\mt^2)$, and
replace all occurrences of
$\ln(\msusyy/\mz^2)$ which multiply either $m_b^4$,
$m_b^2$ or $P_b$ by $\ln(M_Q M_D/\mz^2)$.  
In addition, eq.~(\ref{mthree}) must be
modified by adding the following squared-mass shifts:
\begin{eqnarray} \label{mdelthree}
\Delta\calm_{11}^2 & = & {-g^2 N_c\mzz\cbb\over 32\pi^2\cw2}\Biggl[
(1+4e_b\sw2)\left({m_b^2\over\mzz\cbb}-{1\over 6}\right)
\ln\left(M_Q^2\over M_D^2\right) \nonumber \\
&&\qquad\qquad\qquad\qquad
-{1\over 6}(1-4e_t\sw2)\ln\left(M_Q^2\over M_U^2\right)
\Biggr]\,,\nonumber \\
\Delta\calm_{22}^2 & = & {-g^2 N_c\mzz\sbb\over 32\pi^2\cw2}\Biggl[
(1-4e_t\sw2)\left({m_t^2\over\mzz\sbb}-{1\over 6}\right)
\ln\left(M_Q^2\over M_U^2\right)\nonumber \\
&&\qquad\qquad\qquad\qquad
-{1\over 6}(1+4e_b\sw2)\ln\left(M_Q^2\over M_D^2\right)
\Biggr]\,,\nonumber \\
\Delta\calm_{12}^2 & = & {g^2 N_c\mzz\sb\cb\over 64\pi^2\cw2}\Biggl[
(1-4e_t\sw2)\left({m_t^2\over\mzz\sbb}-{1\over 3}\right)
\ln\left(M_Q^2\over M_U^2\right)\nonumber \\
&&\qquad\qquad\qquad\qquad
+(1+4e_b\sw2)\left({m_b^2\over\mzz\cbb}-{1\over 3}\right)
\ln\left(M_Q^2\over M_D^2\right)\Biggr]\,.
\end{eqnarray}
In the limit of $\mha\gg \mz$ [using eq.~(\ref{largemacase})],
we find that the following squared-mass shift must be added to
$(\mhl^2)_{\rm 1LL}$ given in eq.~(\ref{mhltot}):
\begin{eqnarray}
\Delta\mhl^2 & = & {g^2 N_c\mzz\cos2\beta\over 32\pi^2\cw2}\Bigl[
(1-4e_t\sw2)\left({m_t^2\over\mzz}+{1\over 6}\cos2\beta\right)
\ln\left(M_Q^2\over M_U^2\right)\nonumber \\
&&\qquad\qquad\qquad\qquad
-(1+4e_b\sw2)\left({m_b^2\over\mzz}-{1\over 6}\cos2\beta\right)
\ln\left(M_Q^2\over M_D^2\right)\Bigr]\,.
\label{delmhltot}
\end{eqnarray}

For $m_{H^\pm}$,
replace $\ln(\msusyy/\mt^2)$ in eq.~(\ref{llform}) with
$\half[\ln(M_Q M_U)/\mt^2)+\ln(M_Q M_D)/\mt^2)]$.
In addition, the following squared-mass shift must be added to
$(\mhpm^2)_{\rm 1LL}$ given in eq.~(\ref{llform}):
\begin{eqnarray}
\Delta\mhpm^2 & = & {-g^2 N_c\over 64\pi^2\mw^2}\Biggl\{
{m_t^2 m_b^2\over\sbb\cbb}(M_U^2-M_D^2)^2 g(M_U^2,M_D^2)\nonumber \\
&&\qquad
+{1\over 2}\left[\mw^2\left({m_t^2\over\sbb}+
{m_b^2\over\cbb}\right)-{2\over 3}m_W^4\right]\left[
\ln\left(M_Q^2\over M_U^2\right)+\ln\left(M_Q^2\over
M_D^2\right)\right]\Biggr\}\,,
\label{delmhpm}
\end{eqnarray}
where the function $g(a,b)$ is defined by:
\begin{equation}
g(a,b)\equiv {1\over (a-b)^2}\left[2-{a+b\over a-b}\ln\left({a\over b}
\right)\right]\,.
\label{gdef}
\end{equation}
Note that $g(a,a)=-1/6a^2$, so that in the limit of $M_Q=M_U=M_D$, all
mass shifts given above vanish.

Next, we consider the modifications of the results of Appendix B when
the diagonal soft-supersymmetry-breaking squark
mass parameters are non-degenerate.  The shifts in the squared-mass
matrix elements of the CP-even Higgs boson given in
eq.~(\ref{deltacalms}) are replaced by the following expressions:
\begin{eqnarray} \label{nonudeltacalms}
(\Delta\calmm_{11})_{\rm mix} & = & {g^2N_c\over 16\pi^2\mw^2}\Biggl\{
{m_b^4A_b X_b\over\cbb}\left[2h(M_Q^2,M_D^2)+A_b X_b\, g(M_Q^2,M_D^2)\right]
+{m_t^4\mu^2 X_t^2\over\sbb}\,g(M_Q^2,M_U^2)\nonumber \\
&&\qquad\qquad+\mzz m_t^2\mu\cot\beta\left[X_t\,p_t(M_Q^2,M_U^2)-
\mu\cot\beta\,B(M_Q^2,M_U^2)\right]\nonumber \\
&&\qquad\qquad+\mzz m_b^2A_b\left[X_b\,p_b(M_Q^2,M_D^2)-A_b\,
B(M_Q^2,M_D^2)\right]\Biggr\}\,,\nonumber \\
(\Delta\calmm_{22})_{\rm mix} & = & {g^2N_c\over 16\pi^2\mw^2}\Biggl\{
{m_t^4A_t X_t\over\sbb}\left[2h(M_Q^2,M_U^2)+A_t X_t\,g(M_Q^2,M_U^2)\right]
+{m_b^4\mu^2 X_b^2\over\cbb\,}g(M_Q^2,M_D^2)\nonumber \\
&&\qquad\qquad+\mzz m_b^2\mu\tan\beta\left[X_b\,p_b(M_Q^2,M_D^2)-
\mu\tan\beta\,B(M_Q^2,M_D^2)\right]\nonumber \\
&&\qquad\qquad+\mzz m_t^2A_t\left[X_t\,p_t(M_Q^2,M_U^2)-A_t\,
B(M_Q^2,M_U^2)\right]\Biggr\} \,,\nonumber\\
(\Delta\calmm_{12})_{\rm mix} & = & {-g^2N_c\over 32\pi^2\mw^2}\Biggl\{
{2m_t^4\over\sbb}\mu X_t\left[h(M_Q^2,M_U^2)+A_t X_t\, g(M_Q^2,M_U^2)\right]
\nonumber \\
&&\qquad\qquad
+{2m_b^4\over\cbb}\mu X_b\left[h(M_Q^2,M_D^2)+A_b X_b\,g(M_Q^2,M_D^2)\right]
\nonumber \\
&&\qquad\qquad-\mzz m_b^2\tan\beta\left[(\mu^2+A_b^2)B(M_Q^2,M_D^2)
-X_b Y_b\,p_b(M_Q^2,M_D^2)\right]\nonumber \\
&&\qquad\qquad-\mzz m_t^2\cot\beta\left[(\mu^2+A_t^2)B(M_Q^2,M_U^2)
-X_t Y_t\,p_t(M_Q^2,M_U^2)\right]\Biggr\}\,,
\end{eqnarray}
where the functions $B$, $h$, $p_b$ and $p_t$ are defined as follows:
\begin{eqnarray} \label{functiondefs}
h(a,b)&=&{1\over a-b}\ln\left({a\over b}\right)\,,\nonumber \\
B(a,b)&=&{1\over (a-b)^2}\left[{1\over 2}\left(a+b\right)-{ab\over
a-b}\ln\left({a\over b}\right)\right]\,,\nonumber \\
p_b(a,b)&=&f(a,b)-2e_b\sw2(a-b)g(a,b)\,,\nonumber \\
p_t(a,b)&=&f(a,b)+2e_t\sw2(a-b)g(a,b)\,,
\end{eqnarray}
$g(a,b)$ is given in eq.~(\ref{gdef}) and $f(a,b)$ is given by:
\begin{equation}
f(a,b)={-1\over a-b}\left[1-{b\over a-b}\ln\left({a\over b}\right)\right]\,.
\label{fdef}
\end{equation}
The contributions to $\Delta\calmm_{\rm mix}$ proportional to 
$m_b^4$ or $m_t^4$ were first given in Ref.~\cite{erz}.

To check that eq.~(\ref{nonudeltacalms}) reduces to
eq.~(\ref{deltacalms}) in the limit of $M_Q=M_U=M_D$, one makes use
of $B(a,a)= 1/6a$,
$f(a,a)=-1/2a$, $g(a,a)=-1/6a^2$, and $h(a,a)=1/a$.
In the limit of $\mha\gg\mz$, eq.~(\ref{deltamhs}) is replaced by:
\begin{eqnarray} \label{nonudeltamhs}
(\Delta\mhl^2)_{\rm mix}&=&{g^2 N_c\over 16\pi^2\mw^2}
\Biggl\{m_t^4 X_t^2\left[2h(M_Q^2,M_U^2)+X_t^2\,g(M_Q^2,M_U^2)\right]
\nonumber\\
&&\qquad\qquad m_b^4 X_b^2\left[2h(M_Q^2,M_D^2)+X_b^2\,g(M_Q^2,M_D^2)\right]
\nonumber\\
&&\qquad\qquad +\mzz m_t^2\cos2\beta\left[
(A_t^2-\mu^2\cot^2\beta)B(M_Q^2,M_U^2)
-X_t^2\,p_t(M_Q^2,M_U^2)\right]\nonumber\\
&&\qquad\qquad -\mzz m_b^2\cos2\beta\left[
(A_b^2-\mu^2\tan^2\beta)B(M_Q^2,M_D^2)
-X_b^2\,p_b(M_Q^2,M_D^2)\right]\Biggr\}\,,\nonumber\\
(\Delta\mhh^2)_{\rm mix}&=&{g^2 N_c\over 16\pi^2\mw^2}
\Biggl\{m_t^4 X_tY_t\cot^2\beta\left[2h(M_Q^2,M_U^2)
+X_tY_t\,g(M_Q^2,M_U^2)\right]\nonumber\\
&&\qquad\qquad m_b^4 X_bY_b\tan^2\beta\left[2h(M_Q^2,M_D^2)
+X_bY_b\,g(M_Q^2,M_D^2)\right]\nonumber\\
&&\qquad\qquad +2\mzz m_t^2\cbb\left[X_tY_t\,p_t(M_Q^2,M_U^2)-(\mu^2+A_t^2)
B(M_Q^2,M_U^2)\right]\nonumber\\
&&\qquad\qquad +2\mzz m_b^2\sbb\left[X_bY_b\,p_b(M_Q^2,M_D^2)-(\mu^2+A_b^2)
B(M_Q^2,M_D^2)\right]\Biggr\}\,.
\end{eqnarray}

The shift in the charged Higgs squared-mass given in
eq.~(\ref{deltamhpm}) is replaced by:
\begin{eqnarray}
(\mhpm^2)_{\rm mix}\!\!&=&\!\!{g^2N_c\over 32\pi^2\mw^2}\Biggl\{
 \mu^2\left[{m_t^4\over\sbb}f(M_Q^2,M_U^2)+{m_b^4\over\cbb}f(M_Q^2,M_D^2)
+{2m_t^2m_b^2\over\sbb\cbb}F(M_Q^2,M_U^2,M_D^2)\right]\nonumber\\
&&\qquad\qquad\quad
-{m_b^2m_t^2\over\sbb\cbb}\biggl[A_b^2f(M_Q^2,M_D^2)+
A_t^2f(M_Q^2,M_U^2)+2A_bA_t F(M_Q^2,M_U^2,M_D^2)\nonumber\\
&&\qquad\qquad\qquad\qquad\qquad\qquad +(\mu^2-A_tA_b)^2
G(M_Q^2,M_U^2,M_D^2)\biggr]\nonumber\\
&& -{m_W^2m_t^2\over\sbb}\left(\mu^2\!\left[
f(M_Q^2,M_U^2)+b(M_Q^2,M_U^2)\right]
-\!A_t^2\left[f(M_Q^2,M_U^2)-\!B(M_Q^2,M_U^2)\right]\right)\nonumber\\
&& -{m_W^2m_b^2\over\cbb}\left(\mu^2\!\left[
f(M_Q^2,M_D^2)+b(M_Q^2,M_D^2)\right]
-\!A_b^2\left[f(M_Q^2,M_D^2)-\!B(M_Q^2,M_D^2)\right]\right)\!\Biggr\}\,,
\label{nonudelmhpm}
\end{eqnarray}
where the new functions that appear depend on three variables:
\begin{eqnarray}
G(a,b,c) & = & {f(a,b)-f(a,c)\over b-c}\,,\label{tdef}\\
F(a,b,c) & = & {(b-a)f(a,b)-(c-a)f(a,c)\over b-c}\,.\label{rdef}
\end{eqnarray}
To check that eq.~(\ref{nonudelmhpm}) reduces to eq.~(\ref{deltamhpm})
in the limit of $M_Q=M_U=M_D$, one should note that
$F(a,c,c)=f(c,a)$ and $G(a,c,c)=-g(a,c)$.  To take the limit of $a=c$,
we may use the results quoted below eq.~(\ref{fdef}).

The above results correspond to the leading terms of an
expansion in $(M_1^2-M_2^2)/(M_1^2+M_2^2)$, 
where $M_1^2$, $M_2^2$ are the squared-mass eigenvalues
of the squark mass matrix.  If the squark squared-mass splitting is
large, one needs a slightly better result.  This is
easily obtained from the formulae above for the CP-even Higgs
squared-masses as follows.  Terms multiplying
factors of $m_b$ involve functions of the arguments $M_Q^2$ and
$M_D^2$.  In these terms, replace $M_Q^2$ and $M_D^2$ by the
corresponding bottom squark squared-masses.  Similarly, terms multiplying
factors of $m_t$ involve functions of the arguments $M_Q^2$ and
$M_U^2$.  In these terms, replace $M_Q^2$ and $M_U^2$ by the corresponding
top squark squared-masses.  One can check that this rule correctly
reproduces the terms in the CP-even Higgs squared-mass matrix
proportional to $m_t^4$ and $m_b^4$ obtained by the effective
potential methods of Ref.~\cite{erz}.  Moreover, we have explicitly
verified that the terms proportional to $m_t^2$ are also
correctly reproduced at $\beta=\pi/2$ by comparing with the
exact one-loop computation of Ref.~\cite{hhprl}.   In the case 
of the charged Higgs mass, the improvement required in the case of
large squark squared-mass splitting is more complicated.  In 
particular, the simple rule quoted above does not apply, since 
there are functions appearing in the charged Higgs mass formulae
that contain both $M_U^2$ and $M_D^2$ as arguments, so the
effects of top and bottom squarks do not factorize.  

Finally, we must address the question of effective scales $\mu_q$ and
$\mu_{\tilde q}$ ($q = t,b$)
introduced in Sections 2 and 3, respectively, to account for
the renormalization group improvement. 
By iteratively solving the RGEs with two different
squark masses we can show that the two-loop leading log
corrections can be absorbed into the one-loop expression
by choosing
\beqn
 \ln  {\mu_{t}^2\over m_t^2} = 
{\ln^2(M_{\tilde t_1}/ m_t) + \ln^2(M_{\tilde t_2}/ m_t) \over
 \ln  (M_{\tilde t_1}M_{\tilde t_2}/ m_t^2)}\,,
\eeqn
and similarly for $\mu_{b}$ [with $\mt$ replaced by $\mz$ in the
arguments of the logarithms].  Similarly, one can modify the analysis
at the end of Section 3 to show that the
$\mu_{\tilde q}$ defined below eq.~(\ref{simplemixform}) generalizes 
to $\mu_{\tilde q}\equiv M_{\tilde q_2}$ ($q= t,b)$, {\it i.e.},
the mass of the corresponding heaviest squark mass eigenvalue.

To illustrate the effects of the 
corrections described in this section, we exhibit in Fig.~\ref{hhhfig10} 
the radiatively corrected RG-improved light CP-even Higgs mass as function 
of $\msusy$ for $\mha=A_t=-\mu=M_Q=M_U=M_D=\msusy$ and
two choices of $\tanb=1.5$ and $\tanb=20$.  The graphs have 
been computed using the algorithm of eq.~(\ref{simplemixform}).  The
dotted line is based on the formulae presented in Appendix A and B,
where terms proportional to $\ln(\mha^2/\mzz)$ and the non-leading
logarithmic term proportional to $m_t^2$ have been omitted, as 
described below eq.~(\ref{mhltot}).  The importance of these two terms 
is illustrated by including them in the dashed curve of Fig.~\ref{hhhfig10}.
This curve also includes the effect of replacing $\msusyy$ in the
leading logs of Appendix A by the appropriate product of 
third generation squark masses, as described below eq.~(\ref{tanbrun}).
We call this the ``improved one-scale'' approximation.  Finally, we 
improve on this approximation by employing the correction terms
described in this section, with the replacement of $M_Q$, $M_U$, and $M_D$
by the corresponding third generation squark masses as described below
eq.~(\ref{rdef}).  The result of this improved treatment of third
generation squark thresholds is depicted by the solid line in
Fig.~\ref{hhhfig10}.  One can see that the improved treatment of thresholds
typically introduces no more than a 2 GeV mass shift in the predicted value of
$\mhl$.

\begin{figure}[htbp]
\centerline{\psfig{file=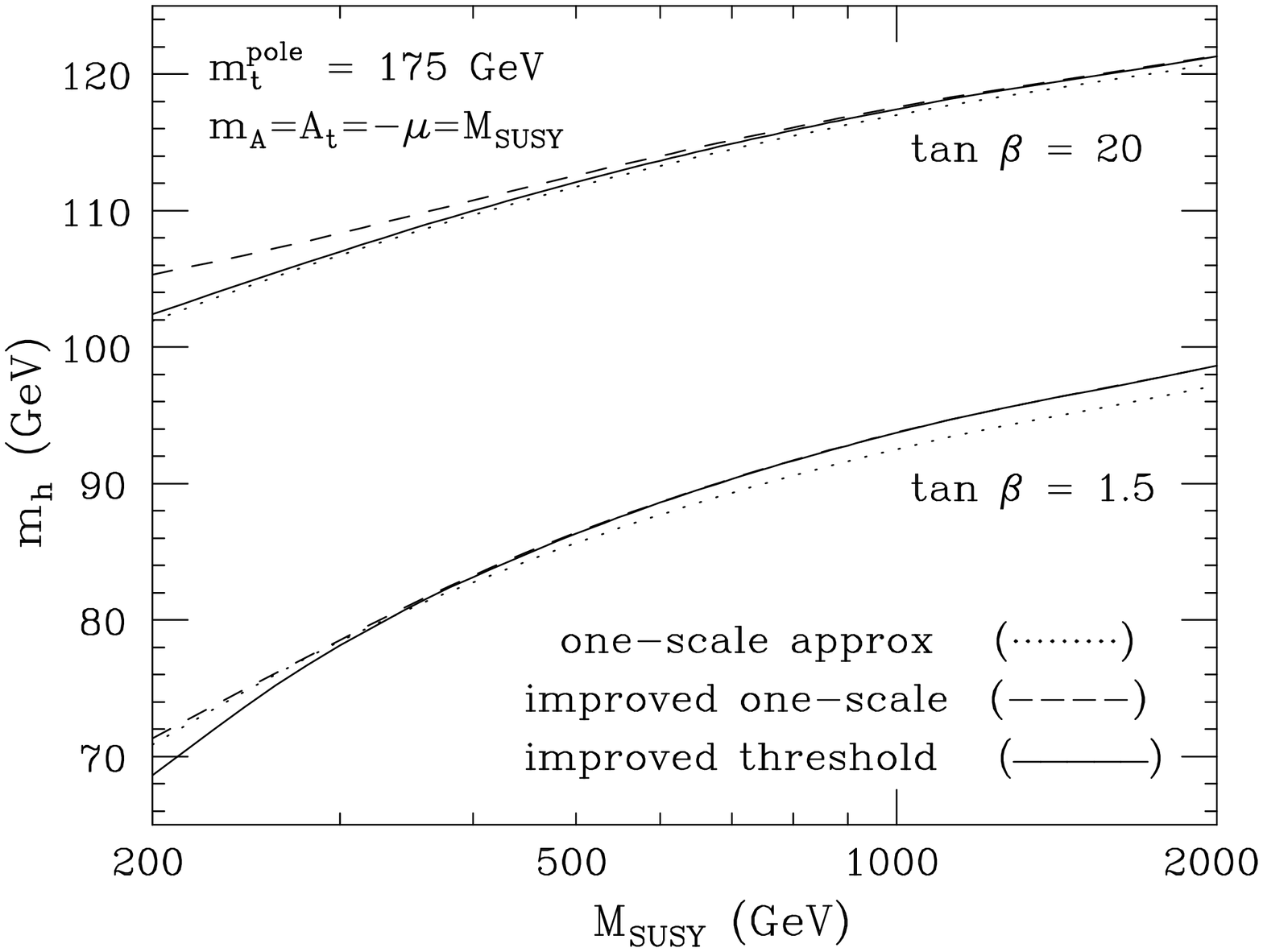,width=12cm,height=9.5cm,angle=90}}
\fcaption{The radiatively corrected light RG-improved 
CP-even Higgs mass is plotted
as a function of $\msusy$ for $\mha= A_t=-\mu=M_Q=M_U=M_D=\msusy$.  
Two choices of
$\tanb=1.5$ and $\tanb=20$ are shown.  The results are based on the
analytic approximation given in eq.~(\ref{simplemixform}).
In the one-scale approximation (dotted line), the distinction between
$\msusy$ and third generation squark masses are neglected.  In addition,
radiative corrections proportional to $\ln(\mha^2/\mzz)$ and the
non-leading log term proportional to $m_t^2$ are neglected, as discussed
below eq.~(\protect\ref{mhltot}).  These approximations were also made
in the graphs of all previous figures.  In the improved one-scale
approximation (dashed line), the latter two neglected terms are included.
In addition, the factors of $\msusyy$ in the argument of the leading logs
of Appendix A are replaced by the appropriate product of 
third generation squark masses, as described below eq.~(\ref{tanbrun}).
In the improved threshold approximation, we include the corrections
of Appendix C with the replacement of $M_Q$, $M_U$, and $M_D$
by the corresponding third generation squark masses as described below
eq.~(\ref{rdef}).}
\label{hhhfig10}
\end{figure}

It is important to emphasize that all the formulae given
in Appendices B and C are based on the assumption that the top and
bottom squark masses are substantially larger than $\mz$.  
Fortunately, in most cases of interest, the error introduced if
the squark masses are not sufficiently heavy is rather small,
of the same size as the non-logarithmic terms that have been
systematically neglected in this paper.
Thus, one can confidently use the formulae given in this paper over
most of the relevant supersymmetric parameter space, and expect an
accuracy in the computed Higgs masses within $\sim 2$~GeV of their actual
values.

%\clearpage
%\par
%\vspace{1cm}\noindent
%{\bf
\appendixD{Appendix D:
Non-universal Gaugino and Higgsino Mass Corrections}

\noindent
The formulae of Appendix A were obtained under the assumption that the
chargino and neutralino masses are all degenerate and equal to
$\msusy$ (assumed large compared to $\mz$).  In this appendix, we relax
the assumption of degeneracy of masses, although we still assume that
all the chargino and neutralino
masses are large compared to $\mz$.  If the latter condition is not
satisfied, then the logarithmic pieces that we keep are of the same order
of magnitude as the non-logarithmic pieces that we omit.  Nevertheless,
the contributions of the charginos and neutralinos to the Higgs masses are
small (never exceeding 2 GeV), so our approximations will be accurate
at this level over the entire neutralino and chargino parameter space.
The results below have been extracted from ref.~\cite{llog}.

In the limit where all chargino and neutralino masses are large compared
to $\mz$, the chargino spectrum consists of two states of mass $M_2$ and
$|\mu|$ respectively, while the neutralino spectrum consists of two
states of mass $M_1$, $M_2$, and two approximately degenerate states of
mass $|\mu|$.  Here, $M_1$ and $M_2$ are Majorana masses for the uncolored
gauginos, and $\mu$ is the supersymmetric Higgs mass
parameter that also appears in
the off-diagonal squark mixing matrix.\footnote{We follow ref.~\cite{pdg2}
for the definitions of the MSSM parameters.}
It will be convenient to introduce
three mass parameters which are equal to either $M_1$, $M_2$, or $|\mu|$
depending on their relative magnitude: 
\beqn \label{threemasses}
&& \mu_1 \equiv {\rm max}\{|\mu|, M_1\} \nonumber\\
&& \mu_2 \equiv {\rm max}\{|\mu|, M_2\} \nonumber\\
&& \mu_{12}\equiv {\rm max}\{|\mu|, M_1, M_2 \}
\eeqn

We consider the case of unequal $M_1$, $M_2$, and $|\mu|$, all of which 
are assumed to be large compared to $\mz$.  Then, eq.~(\ref{mthree}) of
Appendix A is modified by adding the following squared-mass shifts:
\beqn\label{inocontribution}
\Delta\calmm_{11} &=& 
   {g^2\mz^2\cos^2\beta \over 96\pi^2\cw2}\left\{
   6\sw2\left(1-2\sw2\right)\ln\left({\mu_1^2\over\mzz}\right)
   -24\sw2\cw2\ln\left({\mu_{12}^2\over\mzz}\right)\right.\nonumber\\
  &&+6\cw2(3-10\cw2)\ln\left({\mu_2^2\over\mzz}\right)
   -8\cw4\ln\left({M_2^2\over\mzz}\right)\nonumber\\
 && \left.-4(\sw4+\cw4)\ln\left({\mu^2\over\mzz}\right)
  -(P_g+P_{2H})\ln\left({\msusyy\over\mzz}\right)\right\}\,,\nonumber\\
\Delta\calmm_{22} &=& 
   {g^2\mz^2\sin^2\beta \over 96\pi^2\cw2}\left\{
   6\sw2\left(1-2\sw2\right)\ln\left({\mu_1^2\over\mzz}\right)
   -24\sw2\cw2\ln\left({\mu_{12}^2\over\mzz}\right)\right.\nonumber\\
  &&+6\cw2(3-10\cw2)\ln\left({\mu_2^2\over\mzz}\right)
   -8\cw4\ln\left({M_2^2\over\mzz}\right)\nonumber\\
 && \left.-4(\sw4+\cw4)\ln\left({\mu^2\over\mzz}\right)
      -(P_g+P_{2H})\ln\left({\msusyy\over\mzz}\right)\right\}\,,\nonumber\\
\Delta\calmm_{12} &=& 
    {-g^2 \mz^2\sin\beta\cos\beta\over 96\pi^2\cw2}\left\{ 
    6\sw2\left(1+2\sw2\right)\ln\left({\mu_1^2\over\mzz}\right)
   +24\sw2\cw2\ln\left({\mu_{12}^2\over\mzz}\right)\right.\nonumber\\
  &&+6\cw2(3+2\cw2)\ln\left({\mu_2^2\over\mzz}\right)
   -8\cw4\ln\left({M_2^2\over\mzz}\right)\nonumber\\
 && \left.-4(\sw4+\cw4)\ln\left({\mu^2\over\mzz}\right)
      -(P_g^\prime+P_{2H}^\prime)\ln\left({\msusyy\over\mzz}\right)
   \right\}\,,
\eeqn
where the $P_i$ are defined in eq.~(\ref{defpp}).
In the limit of $\mha\gg \mz$ [using eq.~(\ref{largemacase})],
we find that the following squared-mass shift must be added to
$(\mhl^2)_{\rm 1LL}$ given in eq.~(\ref{mhltot}):
\begin{eqnarray}\label{deltaino} 
\Delta\mhl^2 &=&
    {g^2\mz^2\over 96\pi^2\cw2}\left\{
  6\sw2(\cos^2 2\beta-2\sw2)\ln\left({\mu_1^2\over\mzz}\right)
  -24\sw2 \cw2 \ln\left({\mu_{12}^2\over\mzz}\right)\right.\nonumber\\
 && -6\cw2 \left[\cos^2 2\beta(1-4\sw2)+6\cw2\right]
    \ln\left({\mu_2^2\over\mzz}\right) 
    -8\cw4\cos^2 2\beta\ln\left({M_2^2\over\mzz}\right)\nonumber\\
 && -4(\cw4+\sw4)\cos^2 2\beta \ln\left({\mu^2\over\mzz}\right)\nonumber \\
 && \left.   -\left[(P_{g}+P_{2H})(\sbiv+\cbiv)
 -2(P_{g}'+P_{2H}')\sbb\cbb\right]\ln\left({\msusyy\over\mzz}\right)
   \right\}\,.
\end{eqnarray}
Note that the effect of the terms proportional to $\ln(\msusyy/\mzz)$ is to
simply remove the $\msusy$ dependence in eqs.~(\ref{mthree}) and 
(\ref{largemacase}) that arises from the gauge/Higgs/gaugino/higgsino
contributions.  These contributions are now more accurately described
by logarithmic factors that are sensitive to the gaugino/higgsino spectrum.

For $\mhpm^2$, the following squared-mass shift must be added to
$(\mhpm^2)_{\rm 1LL}$ given in eq.~(\ref{llform}):
\begin{eqnarray}
\Delta\mhpm^2 & = & {-g^2 \mw^2 \over 48\pi^2}\left[
3\tan^2\theta_W\ln\left({\mu_1^2\over\mww}\right)
+12\tan^2\theta_W\ln\left({\mu_{12}^2\over\mww}\right)
-3\ln\left({\mu_2^2\over\mww}\right)\right.\nonumber\\
&& \left. -4\ln\left({M_2^2\over\mww}\right)-2\ln\left({\mu^2\over\mww}\right)
-(15\tan^2\theta_W-9)\ln\left({\msusyy\over\mww}\right)\right]\,.
\end{eqnarray}
As above, the effect of the terms proportional to $\ln(\msusyy/\mww)$ is to
simply remove the $\msusy$ dependence in eq.~(\ref{llform})
that arises from the gauge/Higgs/gaugino/higgsino contributions.

One can easily check that if $M_1=M_2=|\mu|=\msusy$, then all the mass 
shifts in this appendix vanish.  
%As the gaugino/higgsino pass parameters
%approach values of order $\mz$, the off-diagonal gaugino-higgsino mixing
%becomes non-negligible, then it becomes a slightly better approximation to
%employ physical chargino and neutralino masses in the arguments of the
%logarithms.  The procedure is similar to that of Appendix C, where we
%replaced $M_Q$, $M_U$ and $M_D$ by physical top squark and bottom squark
%masses.  In the present case, one would replace $M_2$ with the mass of the
%chargino with dominant gaugino component, and $\mu$ with the mass of the 
%chargino with dominant higgsino component.  The masses $\mu_1$, $\mu_2$, and
%$\mu_{12}$ would be replaced by the appropriate neutralino masses 
%({\it e.g.}, $\mu_{12}$ would be replaced by the mass of the heaviest 
%neutralino).     
In the case of the neutral [charged]
Higgs mass computation, if any of the gaugino/higgsino mass parameters is less
than $\mz$ [$\mw$], then one must remove the corresponding logarithmic term
from the above expressions.   
%set the corresponding parameter to $\mz$
%[$\mw$] ({\it i.e.}, the corresponding logarithmic factor vanishes).  
Note that the supersymmetric limit corresponds to all three gaugino/higgsino
mass parameters zero and $\tanb=1$, in which case
two charginos are degenerate with the $W^\pm$ and $\hpm$, 
two neutralinos are degenerate
with the $Z$ and $\hh$, one neutralino is degenerate with the photon,
and one neutralino is degenerate with the (massless) $\hl$ and $\ha$.  In this
case, we must remove all logarithmic terms above, except for the log terms
containing $\msusyy$.  The effect of the latter is to precisely
cancel the contributions of the gauge and Higgs loops to the 
CP-even Higgs squared-mass matrix [eqs.~(\ref{mthree}) and 
(\ref{largemacase})] and the charged Higgs mass [eq.~(\ref{llform})].
%The interpretation of this cancellation is clear.  The effective low-energy
%theory at the scale of $\mz$ [$\mw$] is still supersymmetric; 
%the cancellation of gauge and Higgs loops by gaugino and higgsino loops is
%a consequence of unbroken supersymmetry.  
This cancellation is a consequence of unbroken supersymmetry.
Of course, as emphasized above,
when some of the gaugino/higgsino mass parameters are of order $\mz$, 
then the above corrections terms are small in magnitude and of the same
order as non-logarithmic corrections not included in this analysis.

%%%%%%%%%%%%%%%%%%%%%%%%%%%%%%%%%%%%%%%%%%%%%%%%%%%%%%%%%%%%%%%%%%%%%%%%
%         -------  R  E  F  E  R  E  N  C  E  S  -------
%%%%%%%%%%%%%%%%%%%%%%%%%%%%%%%%%%%%%%%%%%%%%%%%%%%%%%%%%%%%%%%%%%%%%%%%
%%%%%%%%%%%%%%%%%%%%%%%%%%%%%%%%%%%%%%%%%%%%%%%%%%%%%%%%%%%%%%%%%%%%%%%%
\begin{sloppy}
\begin{raggedright}
\def\app#1#2#3{{\sl Act. Phys. Pol. }{\bf B#1} (#2) #3}
\def\apa#1#2#3{{\sl Act. Phys. Austr.}{\bf #1} (#2) #3}
\def\ppnp#1#2#3{{\sl Prog. Part. Nucl. Phys. }{\bf #1} (#2) #3}
\def\npb#1#2#3{{\sl Nucl. Phys. }{\bf B#1} (#2) #3}
\def\jpa#1#2#3{{\sl J. Phys. }{\bf A#1} (#2) #3}
\def\plb#1#2#3{{\sl Phys. Lett. }{\bf B#1} (#2) #3}
\def\prd#1#2#3{{\sl Phys. Rev. }{\bf D#1} (#2) #3}
\def\pR#1#2#3{{\sl Phys. Rev. }{\bf #1} (#2) #3}
\def\prl#1#2#3{{\sl Phys. Rev. Lett. }{\bf #1} (#2) #3}
\def\prc#1#2#3{{\sl Phys. Reports }{\bf #1} (#2) #3}
\def\cpc#1#2#3{{\sl Comp. Phys. Commun. }{\bf #1} (#2) #3}
\def\nim#1#2#3{{\sl Nucl. Inst. Meth. }{\bf #1} (#2) #3}
\def\pr#1#2#3{{\sl Phys. Reports }{\bf #1} (#2) #3}
\def\sovnp#1#2#3{{\sl Sov. J. Nucl. Phys. }{\bf #1} (#2) #3}
\def\jl#1#2#3{{\sl JETP Lett. }{\bf #1} (#2) #3}
\def\jet#1#2#3{{\sl JETP Lett. }{\bf #1} (#2) #3}
\def\zpc#1#2#3{{\sl Z. Phys. }{\bf C#1} (#2) #3}
\def\ptp#1#2#3{{\sl Prog.~Theor.~Phys.~}{\bf #1} (#2) #3}
\def\nca#1#2#3{{\sl Nouvo~Cim.~}{\bf#1A} (#2) #3}
\def\hpa#1#2#3{{\sl Helv.~Phys.~Acta~}{\bf #1} (#2) #3}
\def\aop#1#2#3{{\sl Ann.~of~Phys.~}{\bf #1} (#2) #3}
\def\fP#1#2#3{{\sl Fortschr.~Phys.~}{\bf #1} (#2) #3}
%%%%%%%%%%%%%%%%%%%%%%%%%%%%%%%%%%%%%%%%%%%%%%%%%%%%%%%%%%%%%%%%%%%%%%%%

\end{raggedright}
\end{sloppy}

\begin{thebibliography}{99}
\bibitem{yukrges}
We use the formulae quoted in Appendix A of Ref.~\cite{llog}. Further
citations to the original literature can be found there.
\end{thebibliography}
\end{document}